\documentclass[numberedappendix,appendixfloats]{emulateapj}
\usepackage{url}

\begin{document}

\shorttitle{Panchromatic View of 47 Tuc and the SMC}
\shortauthors{Kalirai et al.}

\title{A Deep, Wide-Field, and Panchromatic View of 47 Tuc and the SMC \\ with HST: Observations and Data Analysis Methods\altaffilmark{1}}

\author{
Jason S.\ Kalirai\altaffilmark{2,3},
Harvey B.\ Richer\altaffilmark{4},
Jay Anderson\altaffilmark{2},
Aaron Dotter\altaffilmark{2},
Gregory G.\ Fahlman\altaffilmark{5}, \\
Brad M.~S.\ Hansen\altaffilmark{6},
Jarrod Hurley\altaffilmark{7},
Ivan R.\ King\altaffilmark{8},
David Reitzel\altaffilmark{6},
R.\ Michael Rich\altaffilmark{6}, \\
Michael M.\ Shara\altaffilmark{9},
Peter B.\ Stetson\altaffilmark{5}, and
Kristin A.\ Woodley\altaffilmark{4}}
\altaffiltext{1}{Based on observations with the NASA/ESA {\it Hubble Space 
Telescope}, obtained at the Space Telescope Science Institute, which is operated 
by the Association of Universities for Research in Astronomy, Inc., under NASA 
contract NAS5-26555.  These observations are associated with proposal GO-11677.}
\altaffiltext{2}{Space Telescope Science Institute, 3700 San Martin Drive, Baltimore, 
MD, 21218; jkalirai@stsci.edu, jayander/dotter@stsci.edu}
\altaffiltext{3}{Center for Astrophysical Sciences, Johns Hopkins University, Baltimore, MD, 21218}
\altaffiltext{4}{Department of Physics \& Astronomy, University of British Columbia, 
Vancouver, BC, Canada; richer@astro.ubc.ca, kwoodley@phas.ubc.ca}
\altaffiltext{5}{National Research Council, Herzberg Institute of Astrophysics, Victoria, BC, 
Canada; greg.fahlman/peter.stetson@nrc-cnrc.gc.ca}
\altaffiltext{6}{Division of Astronomy and Astrophysics, University of California at Los Angeles, 
Los Angeles, CA, 90095; hansen/rmr@astro.ucla.edu, david.reitzel@gmail.com}
\altaffiltext{7}{Center for Astrophysics \& Supercomputing, Swinburne University of Technology, 
Hawthorn VIC 3122, Australia; jhurley@swin.edu.au}
\altaffiltext{8}{Department of Astronomy, Box 351580, University of Washington, Seattle, WA, 98195; 
king@astro.washington.edu}
\altaffiltext{9}{Department of Astrophysics, American Museum of Natural History, Central Park 
West and 79th Street, New York, NY 10024; mshara@amnh.org}

%%%%%%%%%%%%%%%%%%%%%%%%%%%%%%%%%%%%%%%%%%%%%%%%%%%%%%%%%%%%%%%%%%%%%%%%%%%%%%%%%%%%

\begin{abstract}

\noindent In {\it Hubble Space Telescope} ({\it HST}) Cycle 17, we imaged the well known 
globular star cluster 47 Tucanae for 121 orbits using the Wide Field Channel (WFC) 
of the Advanced Camera for Surveys (ACS) and both the UVIS and IR channels of the newly 
installed Wide Field Camera 3 (WFC3) instrument (GO-11677, PI -- H.\ Richer).  This unique 
data set was obtained to address many scientific questions that demand a very deep, panchromatic, 
and panoramic view of the cluster's stellar populations.  In total, the program obtained 
over 0.75 Ms of imaging exposure time with the three {\it HST} cameras, over a time span 
of 9 months in 2010.  The primary ACS field was imaged in the two broadband filters $F606W$ 
and $F814W$ filters, at 13 orientations, for all 121 orbits.  The parallel WFC3 imaging 
provides a panchromatic (0.4 -- 1.7 micron) and contiguous imaging swath over a 250 degree 
azimuthal range at impact radii of 6.5 -- 17.9 pc in 47~Tuc.  This imaging totals over 60 arcmin$^2$ 
in area and utilizes the $F390W$ and $F606W$ broadband filters on WFC3/UVIS 
and the $F110W$ and $F160W$ broadband filters on WFC3/IR.
\\

\noindent In this paper, we describe the observational design of the new survey and one of the methods 
used to analyze all of the imaging data.  This analysis combines over 700 full-frame images taken with 
the three {\it HST} cameras into a handful of ultra-deep, well-sampled combined images in each of the six 
filters.  We discuss in detail the methods used to calculate accurate transformations that provide optimal 
alignment of the input images, the methods used to perform sky background offsets in the input stack and 
the flagging of deviant pixels, and the balance reached between the input pixel drop size onto an output 
super-sampled pixel grid.  Careful photometric, morphological, and astrometric measurements are performed 
on the stacks using iterative PSF-fitting techniques, and reveal unprecedented color-magnitude diagrams (CMDs) 
of the cluster extending to $>$30th magnitude in the optical, 29th magnitude in the UV, and 27th magnitude 
in the IR.  The data set provides a characterization of the complete stellar populations of 47~Tuc, 
extending from the faintest hydrogen burning dwarfs through the main-sequence and giant branches, down to 
very cool white dwarf remnants in the cluster.  The imaging also provides the deepest probe of the stellar 
populations of the background Small Magellanic Cloud (SMC) galaxy, resolving low mass main-sequence dwarfs 
with $M \lesssim$ 0.2~$M_\odot$.

\end{abstract}


%%%%%%%%%%%%%%%%%%%%%%%%%%%%%%%%%%%%%%%%%%%%%%%%%%%%%%%%%%%%%%%%%%%%%%%%%%%%%%%%%%%%

\section{Introduction} \label{introduction}

Observations of the Milky Way's globular star clusters have represented one of the most important 
aspects of astronomy over the past century \citep{shapley17,sandage53}.  Clusters are ideal testbeds for 
theories of stellar structure and evolution as well as dynamical processes, because their constituent 
stars share incredible similarities, being generally co-eval, co-spatial, and iso-metallic 
\citep{harris96,vandenberg85,kalirai10a}.  Photometric measurements of the stars in any given cluster 
therefore provide a snapshot of how stellar and dynamical evolution operate at fixed conditions 
(e.g., age, metallicity, and cluster dynamical state), and this picture can be generalized by observing 
samples of clusters with different conditions.  As one general application, modern day observations of 
nearby globular star clusters provide stringent tests of stellar-evolution models (e.g., Dotter et~al.\ 
2008), which in turn yield a calibration of population-synthesis techniques.  These techniques are 
frequently used to interpret light from distant galaxies into fundamental properties such as star 
formation rates, chemical abundance trends, and mass-to-light ratios (e.g., Bruzual \& Charlot 2003).

In addition to being excellent tools for stellar structure and evolution theory, globular clusters 
also play an important role in our understanding of the epoch of baryonic structure formation 
in the Universe.  Studies of star-forming regions in different environments frequently reveal sites of 
cluster formation (Whitmore \& Schweizer~1995; Hillenbrand 1997; Brodie \& Strader 2006), and therefore much 
observational and theoretical effort has been devoted to deriving accurate absolute ages of these systems.  
The most popular technique for this relies on the comparison of theoretical 
isochrones to the main-sequence turnoff morphology in the color-magnitude diagram (CMD), a feature that 
is dependent on nuclear reaction rates, uncertainties in chemical composition, the equation of state, and 
second-order effects such as diffusion, rotation, and turbulence \citep{renzini88}.  Although the first 
measurements of the turnoff in a nearby globular cluster were made in the 1950s (Sandage 1953), a huge 
leap forward was enabled in the 1980s when CCDs came online and again recently by the high-resolution 
Advanced Camera for Surveys (ACS) on the {\it Hubble Space Telescope} ({\it HST}).  For example, 
\cite{sarajedini07} executed a systematic and homogeneous survey of 65 Milky Way globular clusters and 
resolved the main-sequence turnoff of each of them with high photometric precision \citep{anderson08b}.  
These data have led to a new understanding of the absolute ages and the age-metallicity relation of the 
Milky Way's globular-cluster population, and suggest that the most metal-poor clusters formed $\sim$1~Gyr 
after the Big Bang \citep{hansen07,marinfranch09,dotter10}.

Complementing the many ground- and space-based imaging programs that have studied both specific 
Milky Way globular clusters and the ensemble population, our team has been involved in a decade-long 
campaign to provide ultra-deep imaging of the complete stellar populations in the nearest clusters 
over multiple epochs.  Our first study targeted the nearest globular cluster Messier 4 in 2001 
(Cycle 9; GO-8679) with 123 orbits of {\it HST} observing time with the Wide Field Planetary Camera 
2,\footnote{\cite{bedin09} provide a more recent study of this cluster with ACS.} and the second study 
targeted the second nearest cluster NGC~6397 in 2005 (Cycle 13; GO-10424) with 126 orbits of ACS 
imaging.  These two projects have led to a wealth of scientific investigations that include 1) 
establishing accurate and independent white dwarf cooling ages for the clusters (Hansen et~al.\ 2004; 
2007), 2) constraints on the color-magnitude relation and luminosity and mass function of stars down 
to the lowest-mass dwarfs capable of core hydrogen burning (King et~al.\ 1998; Richer et~al.\ 2004; 
2008), 3) new tests on the internal dynamical state of star clusters (Anderson et~al.\ 2008a; 
Davis et~al.\ 2008a; Hurley et~al.\ 2008), 4) the most accurately constrained external space motions 
around the Milky Way through unprecedented proper motions (Kalirai et~al.\ 2004; 2007; Anderson et~al.\ 
2008a), and 5) a wealth of variability and binary-related studies (Davis et~al.\ 2008b).

Building on the success of these two previous programs, we received 121 orbits of {\it HST} observing 
time in Cycle 17 to establish a detailed imaging survey of one of the most luminous clusters in the 
sky, 47 Tucanae (GO-11677; PI -- H.\ Richer).  47~Tuc is perhaps the best-studied globular cluster (e.g., 
Guhathakurta et~al.\ 1992).  Its rich stellar content has been targeted by large campaigns for studies of 
extrasolar planets (Gilliland et~al.\ 2000), x-ray sources (Heinke et~al.\ 2005), detailed dynamical 
studies (Meylan \& Mayor 1986; McLaughlin et~al.\ 2006), and much more.  The cluster is also the prototype 
metal-rich population in the Galactic halo and provides stringent tests for theories of stellar evolution 
(Demarque \& McClure 1977) and represents a metal-rich anchor for establishing star formation 
histories and abundances of resolved and unresolved stellar population in other galaxies 
(Brown et~al.\ 2003; Bruzual \& Charlot 2003).

\begin{figure*}[ht]
\begin{center}
\leavevmode 
\includegraphics[width=13.0cm, angle=90]{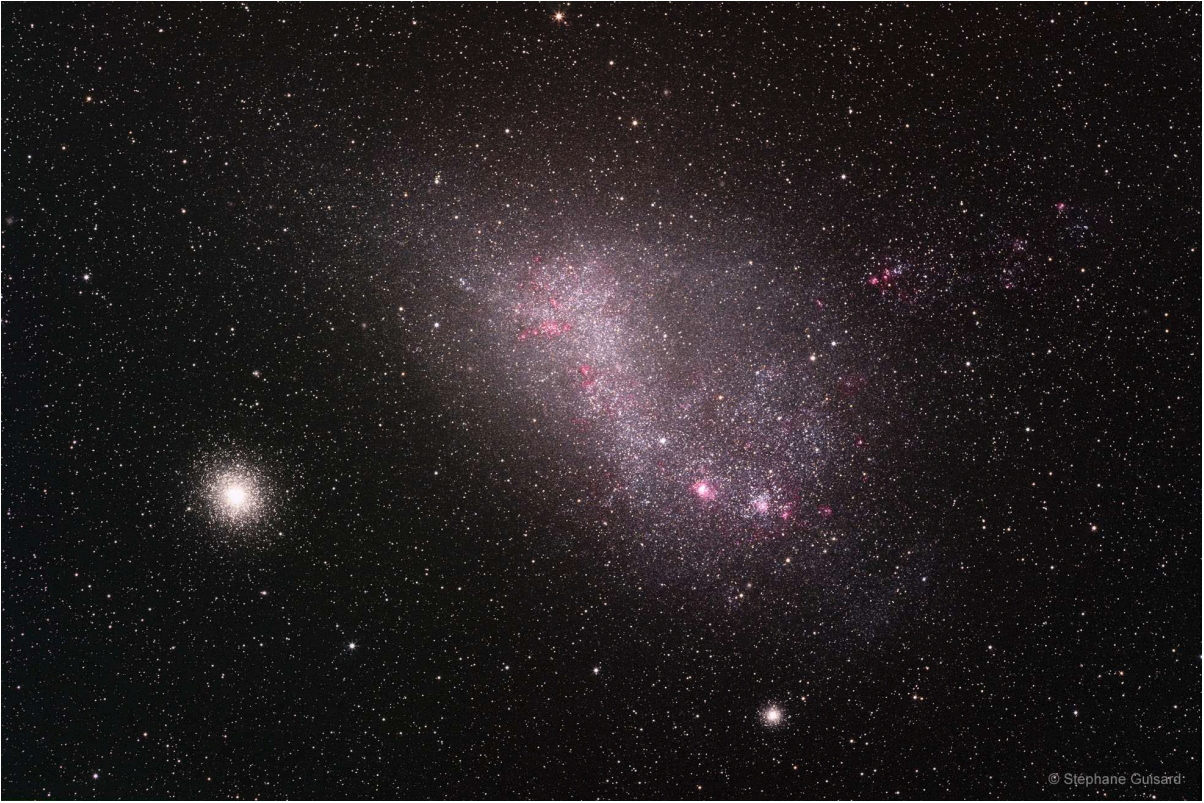}
\end{center}
\vspace{-1.2cm}
\caption{A wide-field ground-based image of the Small Magellanic Cloud (SMC) in the southern skies 
reveals two foreground Milky Way globular clusters, NGC~362 just below the SMC and 47~Tuc to the 
left of the galaxy.  Although the main body of the SMC is separated from 47 Tuc by more than two 
degrees, a diffuse stellar population persists to greater radii and represents a background 
source of stars in our study (as demonstrated later).  This image subtends 6.8 $\times$ 4.5 degrees 
and was taken with a 300~mm lens in September 2007.  The image was made by combining multiple 10 
minute exposures in five visible filters (including H$\alpha$).  Image is courtesy of St\'{e}phane 
Guisard and reproduced here with permission, {\tt http://www.astrosurf.com/sguisard}.
\label{fig:SMCand47Tuc}}
\end{figure*}

Unlike our previous studies of M4 and NGC~6397, this new project utilizes 
three {\it HST} cameras to image the cluster's stellar populations to unprecedented depth over both a 
wide wavelength range ($\lambda$ = 0.4 -- 1.7 microns) and over multiple fields spanning $>$60 arcmin$^2$.  
This data set is aimed at achieving similar science goals to those listed above, yet over this expanded 
region of parameter space (e.g., the calibration of stellar-evolution models in the IR).  The 
observations also yield the deepest probe to date of the lower main sequence of the 
Small Magellanic Cloud (SMC) dwarf galaxy, which is located behind 47 Tuc in projection (see 
Figure~\ref{fig:SMCand47Tuc}).  The primary field is imaged at 13 orientations with 
the ACS/WFC in two visible filters $F606W$ and $F814W$ for all 121 orbits, and the parallel fields 
are mapped out by the newly installed Wide Field Camera 3 instrument using both the UVIS ($F390W$ 
and $F606W$) and IR ($F110W$ and $F160W$) cameras.  The wavelength coverage of these filters is shown 
relative to the spectral energy distribution of three stars that are relevant to the present 
science program in Figure~\ref{fig:spectral}; a hot white dwarf, a moderate temperature globular cluster 
turnoff star, and a cool M dwarf.

%%%%%%%%%%%%%%%%%%%%%%%%%%%%%%%%%%%%%%%%%%%%%%%%%%%%%%%%%%%%%%%%%
%%%%%%%%%%%%%%%%%%%%%%%%%%%%%%%%%%%%%%%%%%%%%%%%%%%%%%%%%%%%%%%%%

%%%%%%%%%%%%%%%%%%%%%%%%%%%%%%%%%%%%%%%%%%%%%%%%%%%%%%%%%%%%%%%%%
%%%%%%%%%%%%%%%%%%%%%%%%%%%%%%%%%%%%%%%%%%%%%%%%%%%%%%%%%%%%%%%%%

%%%%%%%%%%%%%%%%%%%%%%%%%%%%%%%%%%%%%%%%%%%%%%%%%%%%%%%%%%%%%%%%%
%%%%%%%%%%%%%%%%%%%%%%%%%%%%%%%%%%%%%%%%%%%%%%%%%%%%%%%%%%%%%%%%%

\begin{figure*}[ht]
\begin{center}
\leavevmode 
\includegraphics[width=13cm, angle=270]{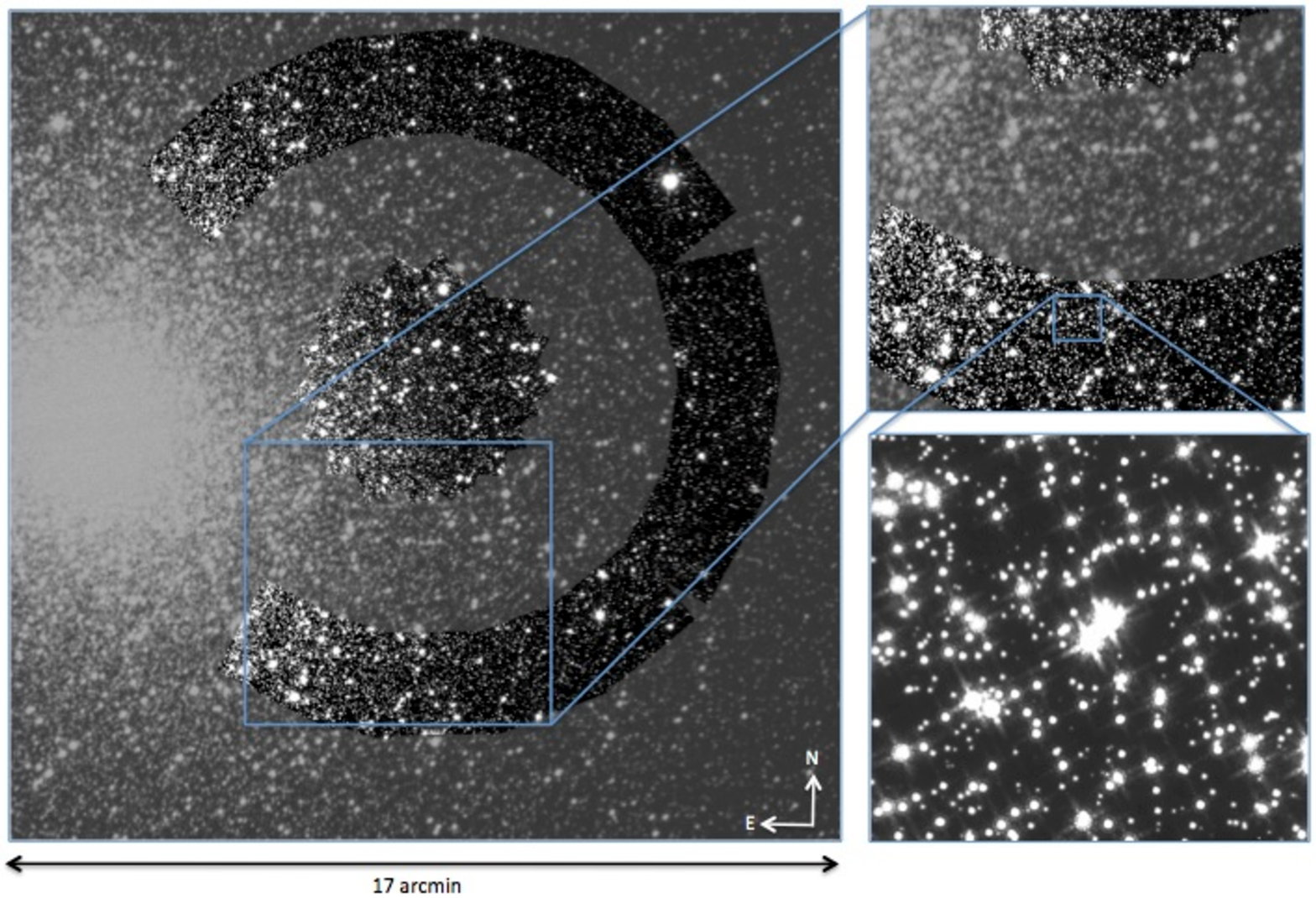}
\end{center}
\vspace{-0.7cm}
\caption{A visualization of the observational strategy which uses three HST cameras.  
The picture in the backdrop is a 
ground based Digital Sky Survey image of 47 Tuc, resampled to a scale of 0.1 arcsec per 
pixel.  The star-like pointing to the west of the cluster center in the center of the mosaic 
represents the deep ACS/WFC field, which was imaged in $F606W$ and $F814W$ for all 121 orbits 
of the program.  The pattern is caused by the requirement to roll the telescope by $\sim$20~degree 
in half of the visits.  These multiple roll angles provide very efficient cleaning of image defects 
and artifacts, and also allow the secondary WFC3 instrument to sweep out an arc around the periphery 
of the cluster.  These parallel fields sample more than 60 arcmin$^2$ of the cluster's population from $R$ = 
6.5 -- 17.9~pc with both the WFC3/UVIS camera in $F390W$ and $F606W$ and the WFC3/IR camera in the 
$F110W$ and $F160W$ filters.  The two panels on the right show zoomed regions of a small 
portion of the parallel WFC3 fields.  This mosaic was constructed by drizzling each pixel in the 
{\it HST} program onto the ground based Digital Sky Survey image, where WFC3/IR $F160W$ was arbitrarily 
selected for the parallel fields.\label{fig:montage}}
\end{figure*}

%%%%%%%%%%%%%%%%%%%%%%%%%%%%%%%%%%%%%%%%%%%%%%%%%%%%%%%%%%%%%%%%%
%%%%%%%%%%%%%%%%%%%%%%%%%%%%%%%%%%%%%%%%%%%%%%%%%%%%%%%%%%%%%%%%%

To support the many on-going scientific projects enabled by this data set, we describe here the detailed 
observational strategy and data-reduction methods that are used to combine over 700 individual full-frame 
images into a single well-sampled stack in each filter which is largely free of cosmetic 
defects.  The methods that are used to measure the photometry, astrometry, and morphology 
of all stellar sources on these combined images are also described.  The final CMDs exhibit well defined 
sequences extending down to 29th magnitude in the UV, 30th magnitude in the visible, and 27th magnitude in the IR, 
and the photometric and astrometric accuracy is assessed using an extensive set of artificial-star tests.  
The primary features of these CMDs are illustrated in relation to the science goals.  The programs 
used for image registration, combination, and photometry in this analysis are all freely available and 
therefore some of the specific techniques used here may be useful for other large {\it HST} programs.

%These observations were motivated by several unique scientific goals that we have studied.  For 
%example, given the old ages of globular clusters, the present day main-sequence turnoff mass is 
%only $\sim$0.8~$M_\odot$ and almost all more massive stars have evolved on to the remnant white 
%dwarf cooling sequence.  Characterizing this sequence yields new insights on the physics of degenerate 
%objects and also provides independent age measurements for the clusters that are not plagued by the 
%theoretical uncertainties listed above (Hansen et~al.\ 2004; 2007).  Our studies also measure the 
%color-magnitude relation of stars extending from bright giants to the faintest main-sequence 
%dwarfs that are capable of burning hydrogen in their cores, and therefore yield stellar luminosity 
%and mass functions over a large range in mass (King et~al.\ 1998; Richer et~al.\ 2004; 2007).  With 
%proper motion measurements (e.g., Anderson et~al.\ 2008a), the observations provide new tests of the 
%internal dynamical state of star clusters (Hurley et~al.\ 2008) and enable unprecedented knowledge 
%on the space motion of the clusters around the Milky Way (Kalirai et~al.\ 2004; 2007).  Finally, 
%the data enable a wide range of variability and binarity-related studies (Davis et~al.\ 2008) that are 
%fundamental to stellar astrophysics.

%%%%%%%%%%%%%%%%%%%%%%%%%%%%%%%%%%%%%%%%%%%%%%%%%%%%%%%%%%%%%%%%%%%

\section{Observational Design} \label{observation}

The observational design of this program was constructed to specifically address several independent 
scientific questions.  The most important consideration was to 
establish a photometric catalog in two filters that provides accurate characterization of the 
faintest white dwarfs in 47~Tuc (i.e., for an independent age measurement).  Based both on white 
dwarf cooling models for low masses and our previous experience from the deep studies of M4 and 
NGC~6397 \citep{hansen04,hansen07}, the white dwarf luminosity function is expected to span over 
5 magnitudes in the CMD and be most efficiently measured using observations at visible wavelengths.   
The secondary 
science goals require mapping the color-magnitude relation and stellar mass function of stars down 
to the faintest dwarfs on the main-sequence.  These stars are cooler than the faintest white 
dwarfs and have redder colors, suggesting observations in the near IR are needed.  A third 
science goal requires establishing a panchromatic data set for the cluster's stellar populations.  
This will enable a sensitive study of the stellar main-sequence and turnoff by stretching out 
features in the CMD and providing enhanced sensitivity to any splittings or multiple sequences, 
and also provide tests of the spectral energy distributions of exotic stellar populations (e.g., 
CVs, blue stragglers, white dwarfs with accretion disks, etc.).  Finally, we require sampling the 
cluster population at a range of radii to test dynamical models of cluster formation and evolution.  

To satisfy these requirements, the design of the program used three HST cameras on two instruments 
operating simultaneously, ACS, WFC3/UVIS, and WFC3/IR.  The primary field was intended to 
reach the faintest magnitudes and requires well-sampled images that are free of image defects caused by 
diffraction spikes, cosmic rays, bad pixels, and charge transfer efficiency (CTE) trails.   The design 
for the parallel fields followed from the primary observations, and achieved greater sensitivity to the 
radial distribution of the stellar population by imaging a larger field of view.  This was achieved by 
rolling the telescope around the primary field multiple times, such that the secondary instrument swept 
out a contiguous arc.  
This design is highlighted in Figure~\ref{fig:montage} through a spatial map of the primary and parallel 
{\it HST} imaging fields superimposed on a ground based image of 47~Tuc.  In the following subsections, 
we first justify and describe the ACS and WFC3 observations in detail, with logs of the exposures 
obtained with all three cameras.  A summary of the overall observational plan is then provided in 
Section~\ref{map}, which also describes this visualization in more detail.

\subsection{The Primary Field -- Advanced Camera for Surveys} \label{acs}

We chose ACS/WFC as the instrument for the deep primary imaging field.  ACS offers a larger field of 
view than WFC3/UVIS and slightly better sensitivity at redder wavelengths.  Observations with ACS can also 
be more easily compared with our previous study of NGC~6397 and 
with previous epoch 47~Tuc images for proper-motion analysis.  The wide-band $F606W$ filter was 
chosen as the primary spectral element given its excellent throughput and a central wavelength that is well 
matched to the peak in the spectral energy distribution of cool white dwarfs.  For the second filter, the 
wide-band $F814W$ spectral element was chosen as it also provides excellent throughput and redder sensitivity 
for secondary science goals related to mapping the low mass main-sequence and red dwarf population of the 
cluster.  

White dwarfs with M = 0.5~$M_\odot$ dim to $M_{F606W}$ $\sim$ 16 after 12~Gyr (Hansen et~al.\ 
2007), translating to a faint observed visual magnitude of $\sim$29.3 at the distance of 47~Tuc 
(4.2~kpc; Zoccali et~al.\ 2001).  We needed 121 orbits of exposure time in a single deep field to 
definitively measure this limit in the white dwarf cooling sequence.  The exposure time was split 
between 117 exposures in $F606W$ and 125 deep exposures in $F814W$.  The exposure times of the $F606W$ 
frames ranged from 1113 -- 1498~seconds for a total integration of 163.7~ks and those in $F814W$ 
ranged from 1031 -- 1484~seconds for an integration 
of 172.8~ks.  Additional exposures of shorter lengths were also obtained to measure bright stars, which 
are saturated on the deep frames.  We obtained 4$\times$100~s, 4$\times$10~s, and 4$\times$1~s exposures 
in $F606W$ and 5$\times$100~s, 4$\times$10~s, and 4$\times$1~s in $F814W$.

Sources on individual {\it HST} images are affected by cosmic rays, hot pixels, CTE losses, diffraction 
spikes from bright saturated stars, possible intrapixel sensitivity variations, and other pixel-to-pixel effects 
from errors in the flat field and geometric distortion.  In order to mitigate these issues, we chose to observe 
the scene at different dither positions and at different roll angles so that a given source would be observed in 
many different pixels at many different places on the detector.  This 
dithering also minimizes errors caused by uncertainties in the pixel-to-pixel flat fielding and 
intra-pixel sensitivities.  The observational strategy yielded two deep images with ACS in each 
orbit, with five orbits being grouped into a visit.  Within each visit, the observations were dithered 
in both the $x$ and $y$ directions.  The $y$ dither ensured that no star would fall in the gap 
between the WFC1 and WFC3 chips on ACS on more than one observation through the same filter.  For visits taken at the 
same orientation, a small $x$ dither ensured that stars would land in different pixels in the different visits.
This dither pattern provides easy discrimination of bad pixels.  Artifacts 
caused by diffraction spikes from the telescope support structure, bleeding of saturated pixels into 
neighboring regions, and CTE losses are more difficult to deal with.  These artifacts are aligned 
with a particular direction and correlated with the position of the source relative to the amplifier 
that reads out the charge, therefore a linear dither pattern will not result in efficient recovery of 
information.  To solve this problem, we constructed the observations such that 13 of the 
24 visits and their corresponding orbits were executed at 13 specific roll angles separated by $\delta$ $\sim$ 
20~degrees.  The locations of stars on the images and their respective read out positions are 
therefore effectively randomized in the field of view.  The remaining 12 fields were imaged at a much 
smaller range in roll angle (within $\pm$5~degree), largely because of {\it HST} scheduling constraints 
and also to ensure a deeper parallel field at fixed location (see below).  An exposure time weighted map 
of these primary field observations is shown in Figure~\ref{fig:context}.

%%%%%%%%%%%%%%%%%%%%%%%%%%%%%%%%%%%%%%%%%%%%%%%%%
%%%%%%%%%%%%%%%%%%%%%%%%%%%%%%%%%%%%%%%%%%%%%%%%%

%%%%%%%%%%%%%%%%%%%%%%%%%%%%%%%%%%%%%%%%%%%%%%%%%%%%%%%%%%%%%%%%%
%%%%%%%%%%%%%%%%%%%%%%%%%%%%%%%%%%%%%%%%%%%%%%%%%%%%%%%%%%%%%%%%%

\begin{figure}[ht]
\begin{center}
\leavevmode 
\includegraphics[width=4.15cm, angle=270, bb = 50 45 400 1084]{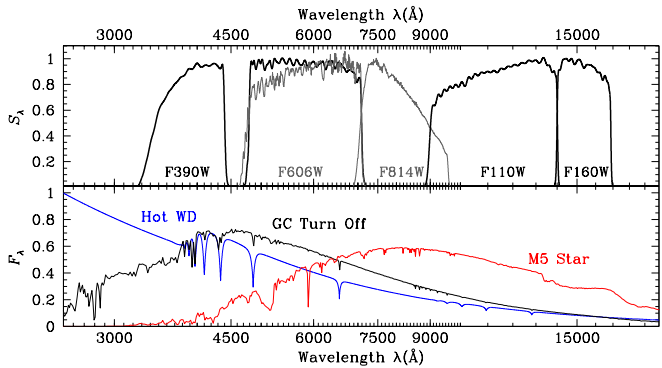}
\end{center}
%\vspace{-1.0cm}
\caption{The wavelength coverage of the ACS (grey), WFC3/UVIS (black), and WFC/IR (black) filters are 
shown relative to the spectral energy distribution of three different stars that are among the target 
populations.  These include a hot white dwarf star (blue), a typical globular cluster turnoff star (black), 
and a cool M dwarf (red).  The WFC3/UVIS $F390W$ filter provides superior throughput compared to bluer 
filters on the instrument, and is ideally suited for 47~Tuc observations given the lack of a hot 
horizontal branch in the cluster.  Note, that the $F606W$ filter was used on both ACS (grey) and WFC3/UVIS 
(black). \label{fig:spectral}}
\end{figure}

%%%%%%%%%%%%%%%%%%%%%%%%%%%%%%%%%%%%%%%%%%%%%%%%%%%%%%%%%%%%%%%%%
%%%%%%%%%%%%%%%%%%%%%%%%%%%%%%%%%%%%%%%%%%%%%%%%%%%%%%%%%%%%%%%%%

%%%%%%%%%%%%%%%%%%%%%%%%%%%%%%%%%%%%%%%%%%%%%%%%%%%%%%%%%%%%%%%%%
%%%%%%%%%%%%%%%%%%%%%%%%%%%%%%%%%%%%%%%%%%%%%%%%%%%%%%%%%%%%%%%%%

\begin{figure}[ht]
%\begin{center}
\leavevmode 
\includegraphics[width=7.5cm, angle=90]{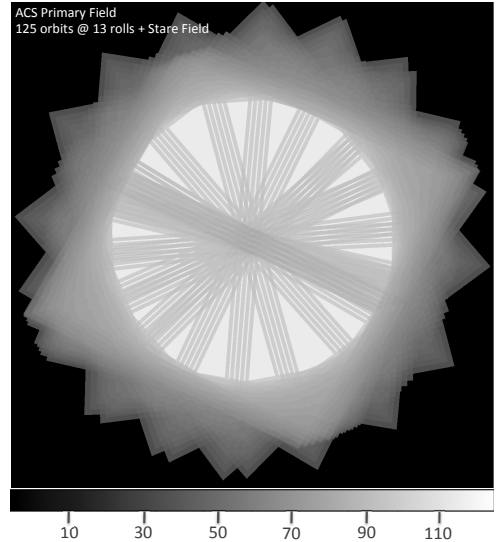}
%\end{center}
\vspace{-0.5cm}
\caption{An exposure time map of the primary ACS field for all 125 $F814W$ images, including the 58 
images at an approximately fixed roll angle of $\sim$293 degrees in the Stare 
field.  The star-like pattern is generated through the multiple roll angles of the observations.  
The color scale gives an indication of the number of images in the stack, where the average exposure time 
of each of the images was 1382~seconds.  The bulk of the field of view has very nice depth of $>$150~ks 
(also in $F606W$), and because of the multiple dithers and roll angles, the actual sky coverage is 
$\sim$50\% larger than the ACS detector's field of view.\label{fig:context}}
\end{figure}

%%%%%%%%%%%%%%%%%%%%%%%%%%%%%%%%%%%%%%%%%%%%%%%%%%%%%%%%%%%%%%%%%
%%%%%%%%%%%%%%%%%%%%%%%%%%%%%%%%%%%%%%%%%%%%%%%%%%%%%%%%%%%%%%%%%

%%%%%%%%%%%%%%%%%%%%%%%%%%%%%%%%%%%%%%%%%%%
%%%%%%%%%%%%%%%%%%%%%%%%%%%%%%%%%%%%%%%%%%%

\begin{deluxetable*}{ccccccccc}
\centering
\small
\tablecaption{The Observational Log of the ACS/WFC Associations$^{1}$\label{ACS.tab}}
\tablecolumns{10}
\tablehead{\colhead{Dataset Name} &
           \colhead{Visit} &
           \colhead{Obs. Date} &
           \colhead{Filter} &
           \colhead{RA} &
           \colhead{DEC} &
           \colhead{Exp.\ Time$^{2}$} &
           \colhead{Orientation} \\
           \colhead{} &
           \colhead{} &
           \colhead{} &
           \colhead{} &
           \colhead{(J2000)} &
           \colhead{(J2000)} &
           \colhead{(s)} &
           \colhead{($^{\circ}$N of E)}} 
\startdata
JB6V01010 & 01 & 2010-02-13 & F606W & 00:22:36.35 & $-$72:04:04.4 & 6920.0 & 316.020 \\
JB6V01020 & 01 & 2010-02-13 & F814W & 00:22:36.35 & $-$72:04:04.4 & 7269.0 & 316.020 \\
JB6V02010 & 02 & 2010-03-04 & F814W & 00:22:36.47 & $-$72:04:00.7 & 6921.0 & 336.126 \\
JB6V02020 & 02 & 2010-03-04 & F606W & 00:22:36.47 & $-$72:04:00.7 & 7323.0 & 336.126 \\
JB6V03010 & 03 & 2010-03-16 & F606W & 00:22:36.80 & $-$72:03:57.8 & 6809.0 & 354.209 \\
JB6V03020 & 03 & 2010-03-16 & F814W & 00:22:36.80 & $-$72:03:57.8 & 7089.0 & 354.209 \\
JB6V04010 & 04 & 2010-04-10 & F814W & 00:22:39.23 & $-$72:04:08.7 & 6930.0 & 17.276  \\
JB6V04020 & 04 & 2010-04-10 & F606W & 00:22:39.23 & $-$72:04:08.7 & 7315.0 & 17.276  \\
JB6V25010 & 25 & 2010-05-03 & F606W & 00:22:38.27 & $-$72:03:53.6 & 6961.0 & 38.312  \\
JB6V25020 & 25 & 2010-05-03 & F814W & 00:22:38.27 & $-$72:03:53.6 & 7269.0 & 38.312  \\
JB6V06010 & 06 & 2010-06-12 & F814W & 00:22:39.07 & $-$72:03:53.7 & 6512.0 & 58.301  \\
JB6V06020 & 06 & 2010-06-12 & F606W & 00:22:39.07 & $-$72:03:53.7 & 7023.0 & 58.301  \\
JB6V07010 & 07 & 2010-06-18 & F606W & 00:22:40.02 & $-$72:03:55.6 & 6920.0 & 84.233  \\
JB6V07020 & 07 & 2010-06-18 & F814W & 00:22:40.02 & $-$72:03:55.6 & 7269.0 & 84.233  \\
JB6V08010 & 08 & 2010-07-29 & F814W & 00:22:40.52 & $-$72:03:57.9 & 8969.0 & 102.157 \\
JB6V08020 & 08 & 2010-07-29 & F606W & 00:22:37.71 & $-$72:04:07.2 & 4125.0 & 102.168 \\
JB6V09010 & 09 & 2010-08-05 & F606W & 00:22:40.84 & $-$72:04:00.9 & 6929.0 & 120.066 \\
JB6V09020 & 09 & 2010-08-05 & F814W & 00:22:40.84 & $-$72:04:00.9 & 7269.0 & 120.066 \\
JB6V10010 & 10 & 2010-08-14 & F814W & 00:22:40.94 & $-$72:04:04.6 & 6930.0 & 139.959 \\
JB6V10020 & 10 & 2010-08-14 & F606W & 00:22:40.94 & $-$72:04:04.6 & 7323.0 & 139.959 \\
JB6V11010 & 11 & 2010-09-19 & F606W & 00:22:40.79 & $-$72:04:08.0 & 6960.0 & 158.861 \\
JB6V11020 & 11 & 2010-09-19 & F814W & 00:22:40.79 & $-$72:04:08.0 & 7269.0 & 158.861 \\
JB6V12010 & 12 & 2010-10-01 & F814W & 00:22:40.40 & $-$72:04:11.0 & 6961.0 & 177.777 \\
JB6V12020 & 12 & 2010-10-01 & F606W & 00:22:40.40 & $-$72:04:11.0 & 7323.0 & 177.777 \\
JB6V13010 & 13 & 2010-01-16 & F606W & 00:22:36.70 & $-$72:04:08.9 & 6791.0 & 287.368 \\
JB6V13020 & 13 & 2010-01-16 & F814W & 00:22:36.70 & $-$72:04:08.9 & 7014.0 & 287.368 \\
JB6V14010 & 14 & 2010-01-17 & F814W & 00:22:36.67 & $-$72:04:09.2 & 6707.0 & 287.368 \\
JB6V14020 & 14 & 2010-01-17 & F606W & 00:22:36.67 & $-$72:04:09.2 & 7153.0 & 287.368 \\
JB6V15010 & 15 & 2010-01-18 & F606W & 00:22:36.66 & $-$72:04:09.5 & 6791.0 & 287.368 \\
JB6V15020 & 15 & 2010-01-18 & F814W & 00:22:36.66 & $-$72:04:09.5 & 7014.0 & 287.368 \\
JB6V16010 & 16 & 2010-01-19 & F814W & 00:22:36.63 & $-$72:04:09.7 & 6707.0 & 287.688 \\
JB6V16020 & 16 & 2010-01-19 & F606W & 00:22:36.63 & $-$72:04:09.7 & 7153.0 & 287.688 \\
JB6V17010 & 17 & 2010-01-20 & F606W & 00:22:36.62 & $-$72:04:08.7 & 6791.0 & 288.829 \\
JB6V17020 & 17 & 2010-01-20 & F814W & 00:22:36.62 & $-$72:04:08.7 & 7014.0 & 288.829 \\
JB6V18010 & 18 & 2010-01-20 & F814W & 00:22:36.57 & $-$72:04:08.8 & 6721.0 & 289.892 \\
JB6V18020 & 18 & 2010-01-21 & F606W & 00:22:36.57 & $-$72:04:08.8 & 7139.0 & 289.892 \\
JB6V19010 & 19 & 2010-01-25 & F606W & 00:22:36.47 & $-$72:04:08.2 & 6791.0 & 294.706 \\
JB6V19020 & 19 & 2010-01-25 & F814W & 00:22:36.47 & $-$72:04:08.2 & 7014.0 & 294.706 \\
JB6V20010 & 20 & 2010-01-23 & F814W & 00:22:36.57 & $-$72:04:09.7 & 9590.0 & 287.871 \\
JB6V20020 & 20 & 2010-01-23 & F606W & 00:22:38.56 & $-$72:04:04.5 & 4270.0 & 287.863 \\
JB6V21010 & 21 & 2010-01-26 & F606W & 00:22:36.62 & $-$72:04:07.7 & 6791.0 & 295.702 \\
JB6V21020 & 21 & 2010-01-26 & F814W & 00:22:36.62 & $-$72:04:07.7 & 7014.0 & 295.702 \\
JB6V22010 & 22 & 2010-01-27 & F814W & 00:22:36.57 & $-$72:04:07.7 & 6707.0 & 296.852 \\
JB6V22020 & 22 & 2010-01-27 & F606W & 00:22:36.57 & $-$72:04:07.7 & 7153.0 & 296.852 \\
JB6V23010 & 23 & 2010-01-28 & F606W & 00:22:36.53 & $-$72:04:07.8 & 6791.0 & 298.002 \\
JB6V23020 & 23 & 2010-01-28 & F814W & 00:22:36.53 & $-$72:04:07.8 & 7014.0 & 298.002 \\
JB6V24010 & 24 & 2010-01-15 & F814W & 00:22:36.69 & $-$72:04:09.8 & 8064.0 & 287.368 \\
JB6V24020 & 24 & 2010-01-15 & F606W & 00:22:36.69 & $-$72:04:09.8 & 8596.0 & 287.368 \\

\enddata
\tablenotetext{1}{The observational log is grouped into {\it HST} visits, where each visit consists 
of five orbits spread over the two filters (except Visit 24 which has six orbits) and each orbit consists 
of two deep exposures and occasional shorter exposures.}
\tablenotetext{2}{The total integration time for all deep and short observations.}
\normalsize
\end{deluxetable*}

%%%%%%%%%%%%%%%%%%%%%%%%%%%%%%%%%%%%%%%%%%%%%%%%%
%%%%%%%%%%%%%%%%%%%%%%%%%%%%%%%%%%%%%%%%%%%%%%%%%

The location of the primary ACS field within 47~Tuc was chosen based on several factors.  The cluster is located 
4.2~kpc from the Sun and spans a large angular size ($r_{t}$ = 42~arcmin = 51~pc).  We require a field that 
is neither too sparse nor too crowded such that an appreciable fraction of faint stars near the termination of 
the white dwarf cooling sequence at $\sim$29th magnitude can be isolated and measured.  We also prefer a field where 
dynamical models suggest that the correction from the observed cluster mass function to the global mass function 
is small, which occurs at $\sim$2 half mass radii according to \cite{hurley08}.  Finally, the field needed to 
overlap previous epoch imaging observations to permit proper-motion separation of the cluster stars from the 
field stars.  All of these requirements were satisfied by centering the ACS primary observations at $\alpha$ = 
00:22:39, $\delta$ = $-$72:04:04, which is located about 6.7 arcmin (8.8~pc) west of the cluster center.  This calibration 
field has been observed more than 245 times in $F606W$ and 66 times in $F814W$ over the lifetime of ACS.

%($r_{h}$ = 3.17~arcmin = 4.1~pc)

\subsection{The Parallel Fields -- Wide Field Camera 3} \label{wfc3}

The newly installed WFC3 instrument provides a nice complement to ACS by offering a second {\it HST} imager 
that operates at high resolution and has a large field of view.  The instrument contains two cameras that are 
optimized for UV/visible wavelengths (UVIS) and for IR wavelengths (IR).  Whereas previous generation HST instruments 
such as NICMOS were frequently used to study distant galaxies, the field of view limited what could be done 
in nearby resolved stellar populations with a large angular extent.  To date, very few deep CMDs with high 
precision photometry exist in the IR bandpasses, and tests of stellar evolution theory in this spectral 
range are almost non-existent.

%%%%%%%%%%%%%%%%%%%%%%%%%%%%%%%%%%%%%%%%%%%%%%%%%
%%%%%%%%%%%%%%%%%%%%%%%%%%%%%%%%%%%%%%%%%%%%%%%%%

\begin{deluxetable*}{ccccccccc}
\centering
\small
\tablecaption{The Observational Log of the WFC3/UVIS Associations$^{1}$\label{WFC3UVIS.tab}}
\tablecolumns{10}
\tablehead{\colhead{Dataset Name} &
           \colhead{Visit} &
           \colhead{Obs. Date} &
           \colhead{Filter} &
           \colhead{RA} &
           \colhead{DEC} &
           \colhead{Exp.\ Time$^{2}$} &
           \colhead{Orientation} \\
           \colhead{} &
           \colhead{} &
           \colhead{} &
           \colhead{} &
           \colhead{(J2000)} &
           \colhead{(J2000)} &
           \colhead{(s)} &
           \colhead{($^{\circ}$N of E)}} 
\startdata
Swath fields \\
IB6V01031 & 01 & 2010-02-13 & F390W & 00:22:34.47 & $-$72:09:58.8 & 2661.0 & 316.020 \\
IB6V01051 & 01 & 2010-02-13 & F606W & 00:22:34.47 & $-$72:09:58.8 & 2849.0 & 316.020 \\
IB6V02031 & 02 & 2010-03-04 & F390W & 00:22:08.32 & $-$72:09:30.7 & 2658.0 & 336.126 \\
IB6V02051 & 02 & 2010-03-04 & F606W & 00:22:08.32 & $-$72:09:30.7 & 2849.0 & 336.126 \\
IB6V03031 & 03 & 2010-03-16 & F390W & 00:21:47.88 & $-$72:08:31.4 & 2606.0 & 354.209 \\
IB6V03051 & 03 & 2010-03-16 & F606W & 00:21:47.88 & $-$72:08:31.4 & 2771.0 & 354.209 \\
IB6V04031 & 04 & 2010-04-10 & F390W & 00:21:29.30 & $-$72:06:38.5 & 2667.0 & 17.283  \\
IB6V04051 & 04 & 2010-04-10 & F606W & 00:21:29.30 & $-$72:06:38.5 & 2845.0 & 17.283  \\
IB6V25031 & 25 & 2010-05-03 & F390W & 00:21:21.98 & $-$72:04:33.5 & 2699.0 & 38.312  \\
IB6V25051 & 25 & 2010-05-03 & F606W & 00:21:21.98 & $-$72:04:33.5 & 2849.0 & 38.312  \\
IB6V06031 & 06 & 2010-06-12 & F390W & 00:21:24.50 & $-$72:02:30.7 & 2548.0 & 58.301  \\
IB6V06051 & 06 & 2010-06-12 & F606W & 00:21:24.50 & $-$72:02:30.7 & 2699.0 & 58.301  \\
IB6V07031 & 07 & 2010-06-18 & F390W & 00:21:40.90 & $-$72:00:10.0 & 2657.0 & 84.233  \\
IB6V07051 & 07 & 2010-06-18 & F606W & 00:21:40.90 & $-$72:00:10.0 & 2849.0 & 84.233  \\
IB6V08031 & 08 & 2010-07-29 & F390W & 00:21:59.33 & $-$71:58:59.0 & 2426.0 & 102.157 \\
IB6V08051 & 08 & 2010-07-29 & F606W & 00:21:59.33 & $-$71:58:59.0 & 2619.0 & 102.157 \\
IB6V09031 & 09 & 2010-08-05 & F390W & 00:22:21.57 & $-$71:58:17.8 & 2666.0 & 120.066 \\
IB6V09051 & 09 & 2010-08-05 & F606W & 00:22:21.57 & $-$71:58:17.8 & 2849.0 & 120.066 \\
IB6V10031 & 10 & 2010-08-14 & F390W & 00:22:48.11 & $-$71:58:11.6 & 2666.0 & 139.959 \\
IB6V10051 & 10 & 2010-08-14 & F606W & 00:22:48.11 & $-$71:58:11.6 & 2849.0 & 139.959 \\
IB6V11031 & 11 & 2010-09-19 & F390W & 00:23:12.33 & $-$71:58:44.9 & 2699.0 & 158.861 \\
IB6V11051 & 11 & 2010-09-19 & F606W & 00:23:12.33 & $-$71:58:44.9 & 2849.0 & 158.861 \\
IB6V12031 & 12 & 2010-10-01 & F390W & 00:23:32.93 & $-$71:59:52.9 & 2699.0 & 177.777 \\
IB6V12051 & 12 & 2010-10-01 & F606W & 00:23:32.93 & $-$71:59:52.9 & 2849.0 & 177.777 \\
\\
Stare field \\
IB6V13031 & 13 & 2010-01-16 & F390W & 00:23:11.84 & $-$72:09:24.1 & 2614.0 & 287.368 \\
IB6V13051 & 13 & 2010-01-16 & F606W & 00:23:11.84 & $-$72:09:24.1 & 2764.0 & 287.368 \\
\enddata
\tablenotetext{1}{The observational log is grouped into {\it HST} visits, where each visit with WFC3/UVIS 
consists of two orbits (one per filter) and each orbit consists of two deep and two short exposures.}
\tablenotetext{2}{The total integration time for all deep and short observations.}
\normalsize
%\end{center}
\end{deluxetable*}

%%%%%%%%%%%%%%%%%%%%%%%%%%%%%%%%%%%%%%%%%%%%%%%%%%%%%%
%%%%%%%%%%%%%%%%%%%%%%%%%%%%%%%%%%%%%%%%%%%%%%%%%%%%%%

Several aspects of the parallel WFC3 observations were automatically set by the primary ACS design.  The 
angular separation of the ACS and WFC3 instruments is $\sim$6$'$, so the 13 different roll angles 
of the primary field map to 13 separate parallel pointings.  The $\sim$20~degree step size in the roll 
angle between each of the primary field observations also ensured that the parallel fields form a contiguous 
imaging region with the WFC3/IR camera (smaller field of view than WFC3/UVIS -- 4.6 arcmin$^2$).  The telescope 
orientation was chosen such that the parallel fields swept out an arc to the 
west of the primary pointing, away from the dense region near the cluster center.  This arc extends 250 
degrees in azimuth and over a radial range of 5 -- 13.7 arcmin  (6.5 -- 17.9~pc), thereby sampling the cluster 
population over a large dynamical range (47 Tuc's core radius is 0.36~arcmin).  The total area of the cluster 
imaged in these pointings is $>$60 sq arcmin with WFC3/IR (larger with WFC3/UVIS).  In referring to 
the WFC3 parallel fields from here on, the deep field with the $\pm$5~degree fixed roll angle are labelled as the 
``Stare'' field (observed for 61 orbits) and the 12 other fields as the ``Swath'' fields (observed for 5 orbits 
each).

%%%%%%%%%%%%%%%%%%%%%%%%%%%%%%%%%%%%%%%%%%%%%%%%%
%%%%%%%%%%%%%%%%%%%%%%%%%%%%%%%%%%%%%%%%%%%%%%%%%

\begin{deluxetable*}{ccccccccc}
\centering
\small
\tablecaption{The Observational Log of the WFC3/IR Associations$^{1}$\label{WFC3IR.tab}}
\tablecolumns{10}
\tablehead{\colhead{Dataset Name} &
           \colhead{Visit} &
           \colhead{Obs. Date} &
           \colhead{Filter} &
           \colhead{RA} &
           \colhead{DEC} &
           \colhead{Exp.\ Time$^{2}$} &
           \colhead{Orientation} \\
           \colhead{} &
           \colhead{} &
           \colhead{} &
           \colhead{} &
           \colhead{(J2000)} &
           \colhead{(J2000)} &
           \colhead{(s)} &
           \colhead{($^{\circ}$N of E)}} 
\startdata
Swath fields \\
IB6V01070 & 01 & 2010-02-13 & F110W & 00:22:35.86 & $-$72:10:00.5 & 3075.64 &  316.006 \\
IB6V01080 & 01 & 2010-02-13 & F160W & 00:22:35.87 & $-$72:10:00.5 & 5993.86 &  316.006 \\
IB6V02070 & 02 & 2010-03-04 & F110W & 00:22:09.51 & $-$72:09:34.5 & 3075.64 &  336.113 \\
IB6V02080 & 02 & 2010-03-04 & F160W & 00:22:08.42 & $-$72:09:32.8 & 5993.86 &  336.117 \\
IB6V03070 & 03 & 2010-03-16 & F110W & 00:21:48.75 & $-$72:08:36.7 & 3035.64 &  354.198 \\
IB6V03080 & 03 & 2010-03-16 & F160W & 00:21:48.75 & $-$72:08:36.7 & 5943.86 &  354.198 \\
IB6V04070 & 04 & 2010-04-10 & F110W & 00:21:29.65 & $-$72:06:44.9 & 3075.64 &  17.276  \\
IB6V04080 & 04 & 2010-04-10 & F160W & 00:21:29.65 & $-$72:06:44.9 & 5993.86 &  17.276  \\
IB6V25070 & 25 & 2010-05-03 & F110W & 00:21:21.81 & $-$72:04:40.1 & 3075.64 &  38.310  \\
IB6V25080 & 25 & 2010-05-03 & F160W & 00:21:21.81 & $-$72:04:40.1 & 5993.86 &  38.310  \\
IB6V06070 & 06 & 2010-06-12 & F110W & 00:21:23.84 & $-$72:02:36.6 & 2972.70 &  58.303  \\
IB6V06080 & 06 & 2010-06-12 & F160W & 00:21:23.84 & $-$72:02:36.6 & 5693.86 &  58.303  \\
IB6V07070 & 07 & 2010-06-18 & F110W & 00:21:39.75 & $-$72:00:14.0 & 3075.64 &  84.241  \\
IB6V07080 & 07 & 2010-06-18 & F160W & 00:21:39.75 & $-$72:00:14.0 & 5993.86 &  84.241  \\
IB6V08070 & 08 & 2010-07-29 & F110W & 00:21:57.98 & $-$71:59:01.2 & 2871.93 &  102.168 \\
IB6V08090 & 08 & 2010-07-29 & F160W & 00:21:57.98 & $-$71:59:01.2 & 5668.86 &  102.168 \\
IB6V09070 & 09 & 2010-08-05 & F110W & 00:22:20.14 & $-$71:58:17.9 & 3075.64 &  120.079 \\
IB6V09080 & 09 & 2010-08-05 & F160W & 00:22:20.14 & $-$71:58:17.9 & 5993.86 &  120.079 \\
IB6V10070 & 10 & 2010-08-14 & F110W & 00:22:46.75 & $-$71:58:09.5 & 3075.64 &  139.972 \\
IB6V10080 & 10 & 2010-08-14 & F160W & 00:22:46.75 & $-$71:58:09.5 & 5993.86 &  139.972 \\
IB6V11070 & 11 & 2010-09-19 & F110W & 00:23:11.19 & $-$71:58:40.9 & 3075.64 &  158.873 \\
IB6V11080 & 11 & 2010-09-19 & F160W & 00:23:11.19 & $-$71:58:40.9 & 5993.86 &  158.873 \\
IB6V12070 & 12 & 2010-10-01 & F110W & 00:23:32.14 & $-$71:59:47.4 & 3075.64 &  177.788 \\
IB6V12080 & 12 & 2010-10-01 & F160W & 00:23:32.14 & $-$71:59:47.4 & 5993.86 &  177.788 \\
\\                                                                      
Stare field \\
IB6V13070 & 13 & 2010-01-16 & F110W & 00:23:12.24 & $-$72:09:25.4 & 3004.95 &  287.360 \\
IB6V13080 & 13 & 2010-01-16 & F160W & 00:23:12.24 & $-$72:09:25.4 & 5743.87 &  287.360 \\
IB6V14040 & 14 & 2010-01-17 & F110W & 00:23:11.22 & $-$72:09:28.3 & 4404.19 &  287.364 \\
IB6V14030 & 14 & 2010-01-17 & F160W & 00:23:10.23 & $-$72:09:30.9 & 9941.56 &  287.368 \\
IB6V15030 & 15 & 2010-01-18 & F110W & 00:23:10.22 & $-$72:09:31.2 & 5785.59 &  287.368 \\
IB6V15050 & 15 & 2010-01-18 & F160W & 00:23:12.20 & $-$72:09:25.9 & 8748.82 &  287.360 \\
IB6V16040 & 16 & 2010-01-19 & F110W & 00:23:11.79 & $-$72:09:27.1 & 4378.42 &  287.680 \\
IB6V16030 & 16 & 2010-01-19 & F160W & 00:23:09.80 & $-$72:09:32.3 & 9967.33 &  287.688 \\
IB6V17040 & 17 & 2010-01-20 & F110W & 00:23:10.40 & $-$72:09:29.1 & 4378.42 &  288.821 \\
IB6V17030 & 17 & 2010-01-20 & F160W & 00:23:08.39 & $-$72:09:34.2 & 9967.33 &  288.829 \\
IB6V18030 & 18 & 2010-01-20 & F110W & 00:23:07.03 & $-$72:09:37.0 & 5596.93 &  289.892 \\
IB6V18040 & 18 & 2010-01-21 & F160W & 00:23:09.07 & $-$72:09:32.1 & 8748.82 &  289.884 \\
IB6V19040 & 19 & 2010-01-25 & F110W & 00:23:02.74 & $-$72:09:42.5 & 4378.42 &  294.698 \\
IB6V19030 & 19 & 2010-01-25 & F160W & 00:23:00.62 & $-$72:09:46.6 & 9967.33 &  294.706 \\
IB6V20040 & 20 & 2010-01-23 & F110W & 00:23:11.52 & $-$72:09:27.5 & 4378.42 &  287.863 \\
IB6V20030 & 20 & 2010-01-23 & F160W & 00:23:09.52 & $-$72:09:32.7 & 9967.33 &  287.871 \\
IB6V21040 & 21 & 2010-01-26 & F110W & 00:23:01.62 & $-$72:09:44.0 & 4503.42 &  295.693 \\
IB6V21030 & 21 & 2010-01-26 & F160W & 00:22:59.49 & $-$72:09:47.9 & 9842.33 &  295.702 \\
IB6V22050 & 22 & 2010-01-27 & F110W & 00:23:00.11 & $-$72:09:46.3 & 4271.16 &  296.843 \\
IB6V22030 & 22 & 2010-01-27 & F160W & 00:22:57.96 & $-$72:09:50.0 & 10074.6 &  296.852 \\
IB6V23040 & 23 & 2010-01-28 & F110W & 00:22:58.59 & $-$72:09:48.5 & 4471.16 &  297.993 \\
IB6V23030 & 23 & 2010-01-28 & F160W & 00:22:56.42 & $-$72:09:52.0 & 9842.33 &  298.002 \\
IB6V24040 & 24 & 2010-01-15 & F110W & 00:23:10.25 & $-$72:09:31.6 & 5670.40 &  287.368 \\
IB6V24030 & 24 & 2010-01-15 & F160W & 00:23:10.25 & $-$72:09:31.6 & 11484.8 &  287.368 \\
\enddata
\tablenotetext{1}{The observational log is grouped into {\it HST} visits, where each visit with WFC3/IR 
consists of three orbits in the Swath field (one for $F110W$ and two for $F160W$) and about five orbits in 
the Stare field for both filters.  Multiple deep and shallow exposures are obtained in each orbit.  
For the Swath field, these are the same visits as those shown in Table~2 for the WFC3/UVIS observations.}
\tablenotetext{2}{The total integration time for all deep and short observations.}
\normalsize
%\end{center}
\end{deluxetable*}

%%%%%%%%%%%%%%%%%%%%%%%%%%%%%%%%%%%%%%%%%%%%%%%%%%%%%%
%%%%%%%%%%%%%%%%%%%%%%%%%%%%%%%%%%%%%%%%%%%%%%%%%%%%%%

The science goals in the program require imaging the stellar populations in the Swath and 
Stare fields both at UV and IR wavelengths.  The separate data sets in each of these regimes 
are very valuable.  For example, the UV observations over this wide field of view will provide 
important constraints on the luminosity function of the brightest white dwarfs that cool via 
neutrino emission, and also provide deep imaging for counterpart studies to exotic stellar 
populations such as LMXBs and CVs.  The IR data are intended to completely characterize the 
47~Tuc main-sequence from the brightest giants to the reddest dwarfs near the hydrogen burning 
limit.  Taken together, this imaging also provides matched catalogs of stars measured over the 
full wavelength baseline from UV through the optical to IR.

The WFC3 observations were split over four wide-band filters.  For the UVIS camera, the $F390W$ 
and $F606W$ filters were chosen, and two long dithered exposures ($\gtrsim$1200~s) were obtained 
in a single orbit for each of the two filters.  The $F390W$ filter was picked over bluer UV filters on WFC3 for two main 
reasons.  First, 47~Tuc is a metal-rich globular cluster with [Fe/H] = $-$0.70 \citep{carretta09} and does 
not contain an extended blue horizontal branch.  Second, the exposure times in each WFC3/UVIS field were fairly 
short and therefore we prefered $F390W$ for its substantially better throughput compared to $F336W$, 
$F275W$, and other UV filters.  The choice of $F606W$ was made to ensure a high throughput 
in these shallow observations and to provide a visible band measurement for the panchromatic study.  
To characterize the brighter stars in the UV data, two dithered 50~s exposures in each of the $F390W$ and 
$F606W$ filters were also obtained, for each of the 13 fields.  The total exposure time with 
WFC3/UVIS in the program was 34.3~ks for the 52 $F390W$ images and 36.5~ks for the 52 $F606W$ images 
(only 1300~s total came from the shallower 50~s exposures in each filter).  As demonstrated later, 
a single orbit's observations in each of these filters (in each of the fields) with WFC3/UVIS provides 
characterization of 47~Tuc stars down to $>$28th magnitude.

For the IR camera, the two widely used wide-band filters $F110W$ and $F160W$ were picked for their high 
throughput and sampling of the full available IR wavelength range.  In each of the 12 Swath fields, 
two dithered exposures with $F110W$ of length 1200 -- 1400~s were obtained in a single orbit (total integration of 
33.3~ks for the 24 images) and four dithered exposures with $F160W$ of length 1200 -- 1400~s were obtained 
in two orbits (total integration of 57.6~ks for the 48 images).  An equal number of images with shorter exposure times  
of 100 -- 350~s were also obtained in each field (3.2~ks in $F110W$ and 13.7~ks in $F160W$).  Unlike for the 
WFC3/UVIS observations, the Stare field was treated differently for WFC3/IR by investing a substantial 
integration to achieve a deeper, well sampled image in each IR filter.  Altogether, 38 images in 
$F110W$ were collected with individual integration times of 1200 -- 1400~s (51.2~ks total) and 80 images in 
$F160W$ were collected with integration times of 1200 -- 1400~s (103.9~ks total).  Shallower exposures of 32 
-- 300~s were also obtained and total 4.1~ks in $F110W$ and 10.4~ks in $F160W$.  The combined Stare field observations 
with WFC3/IR total 59 orbits alone.

\subsection{Summary of Observations} \label{map}

Given the complex design with multiple roll-angle restrictions, the observations described above 
were collected over multiple epochs extending from January to October 2010.  
In total we obtained 708 full frame images with ACS, WFC3/UVIS, and WFC3/IR with 
a total exposure time of 0.75 Ms.  In the next section, we discuss an analysis of these data 
that reduces the complete data set to a set of single images in each filter for the deep and 
shallow observations.  

In Figure~\ref{fig:montage}, a visualization of these {\it HST} data was presented.   This figure illustrates 
a montage of the ACS primary field and the WFC3/IR Stare and Swath fields, superimposed within a 
ground-based Digital Sky Survey image of 47 Tuc.  The ground-based image in the background has 
been resampled to 0.1 arcsec / pixel and subtends an angular scale of $\sim$16.7~arcmin 
(i.e., 10,000 pixels $\times$ 10,000 pixels).  The positions and fluxes of all 
sources on this image were measured to produce a catalog, and astrometrically aligned using 2MASS standards.
This master astrometric grid was first used to calculate transformations between each of the new 
{\it HST} images and the catalog, and then every pixel of every deep image that we obtained 
with ACS and WFC3/IR was drizzled on to that grid (i.e., the {\it HST} images were also rescaled 
to 0.10~arcsec).  The intensity of this master {\it HST} montage was scaled to match the ground-based 
image, the two were added together, and we retained pixels from the {\it HST} image only where they 
were available and those of the ground-based image otherwise.

The montage shows the primary ACS field from our study just west of the cluster center as a 
star-like pattern.  This results from the multiple roll angles, each at $\delta$ $\sim$ 20~degrees.  
Surrounding this image in an arc are the multiple fields observed with WFC3, beginning at the bottom 
just south-east of the ACS field (the Stare field) and continuing to the north in the 12 separate 
Swath fields.  This arc is illustrated here with the WFC3/IR imaging fields, whereas, the WFC3/UVIS 
coverage is much better.

A summary of all of the observations collected in this program is presented in Table~1 (ACS), 
Table~2 (WFC3/UVIS), and Table~3 (WFC3/IR).  In these tables, all exposures that are obtained within 
single visits are grouped together, where the visits include multiple dithered exposures 
over several orbits.  The identifier in column 1 represents the association name for the first 
exposure in each group.  More detailed information on the 707 individual exposures can be 
obtained from the Multimission Archive at STScI (MAST -- {\tt http://archive.stsci.edu}).

%Harris 47 Tuc - Core Radius = 0.36'
%              - Half Light Radius = 3.17'

%############

%ACS  CCD gap is 50 pixels
%WFC3 CCD gap is 35 pixels
%WFC3/UVIS FWHM = 1.6-2.3 pixels

%############

%117 ACS V Deep
%12 ACS V Bright
%
%125 ACS I Deep
%12  ACS I Bright
%
%38 J Stare Deep
%19 J Stare Inter
%9  J Stare Bright
%
%80 H Stare Deep
%21 H Stare Inter
%27 H Stare Bright
%
%24 J Swath Deep
%24 J Swath Inter
%
%48 H Swath Deep
%48 H Swath Inter
%
%26 C Swath Deep
%26 C Swath Inter
%
%26 V Swath Deep
%26 C Swath Deep

%##############

%F110W Swath F1-F12 = 33.3 ks, 1199-1399 s exposure times all with STEP200 and 14-15 NSAMPs -- 24 images
%F110W Swatch F1-F12 = 3.2~ks, 100-174 s exposure time with STEP25 and SPARS10 8-12 NSAMPS  -- 23 images

%F160W Swath F1-F12 = 57.6 ks, 1199 s exposure times with STEP400 and 12 NSAMPs  - 48 images
%F160W Swath F1-F12 = 13.7 ks, 100-350 s exposure times with STEP50 and 9-13 NSAMPS -- 48 images

%F110W Stare = 51.2 ks, 1199-1399 s exposure times with STEP200 and STEP400 and NSAMP = 12-15 - 38 images
%F110W Stare = 4.1 ks, 32-299 s exposure times with range of outputs and 12-15 NSAMPS - 28 images

%F160W Stare = 103.9 ks, 1199-1399 s exposure times with STEP200 and STEP400 and 12-15 NSAMP = 80 images
%F160W Stare = 10.4 ks, 32-299 s exposure times with range of outputs and 12-15 NSAMPS - 48 images

%%%%%%%%%%%%%%%%%%%%%%%%%%%%%%%%%%%%%%%%%%%%%%%%%%%%%%%%%%%%%%%%%%%

\section{Image Analysis and Stacking} \label{datareduction}

The first images of this program were collected by {\it HST} in January 2010, just 4 months after 
STScI finished the Servicing Mission Orbital Verification (SMOV) program that provided a preliminary 
calibration of the repaired ACS and newly installed WFC3 instruments.  For ACS, post SM4 calibration 
has shown that the instrument's behavior, including photometric throughput and anomalies, is similar 
to the existing baseline that was established before it shut down in January 2007.  For WFC3, however, 
the learning curve has initially been very steep.  Calibration reference files were produced with the 
first on-orbit data and these replaced the ground-calibration files.  New on-orbit files have subsequently 
been released through the web pages and updated into the data reduction pipeline system as new knowledge 
of the respective instrument performance and defects has become available. 

\begin{figure}[ht]
\begin{center}
\leavevmode 
\includegraphics[width=8.8cm, angle=270]{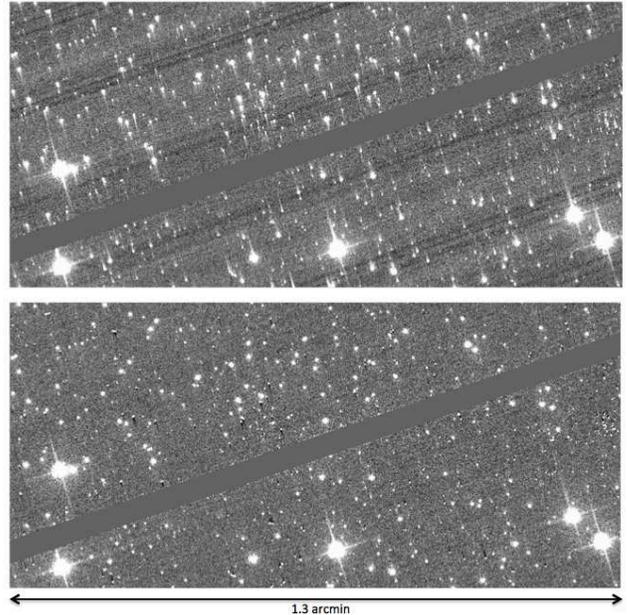}
\end{center}
\vspace{-0.5cm}
\caption{{\it Top} -- An individual 10~s \_flt image with a background of $\sim$1~$e^{-}$ from the ACS/WFC 
data shows significant effects of CTE losses.  The trails are caused by electrons that were released after a 
short delay, being trapped due to pixel degradation of the detectors.  The objects that are nearest 
to the chip gap shown are the furthest from the readout registers at the top and bottom of the detector, and 
therefore encounter the most charge traps and suffer the strongest trails.  {\it Bottom} -- The same image 
corrected for CTE losses using the methods of Anderson \& Bedin (2010).  Both the flux and position of sources is 
restored.  Note, the image in the top panel shows faint streaks that are parallel to the chip gap.  These are 
present in post-SM4 ACS data, and are caused by readout electronics.  Although the streaks look bad, their 
amplitude is very low.  In the bottom panel, we remove these streaks using the acs\_destripe Python task \citep{grogin10}.
\label{fig:CTE}}
\end{figure}

Given the frequency of these updates for WFC3, the data from the two instruments in this program were treated 
differently.  For ACS, the calibrated \_flt files which were produced using the {\it calacs} pipeline with 
post-SM4 reference files were retrieved from MAST. Recent ACS images exhibit clear signatures of CTE 
degradation.  Traps in the silicon lattice cause some charge to be delayed during readout, essentially 
blurring the profiles of sources away from the readout amplifier.  Charge from a source is often delayed 
so much that it is recorded well outside of the aperture used to measure the object, with the result 
being that the object appears fainter than it truly is (see Figure~\ref{fig:CTE}).  Not correcting for this 
effect would lead to both larger photometric and astrometric errors in the data, especially on short 
exposures with low sky background \citep{anderson10,massey10}.  All ACS pipeline \_flt images were 
therefore subjected to the pixel based CTE correction task PixCteCorr in Python, which is described in detail on the 
STScI ACS webpages\footnote{\tt http://www.stsci.edu/hst/acs/software/CTE/} and \cite{anderson10}.

%%%%%%%%%%%%%%%%%%%%%%%%%%%%%%%%%%%%%%%%%%%%%%%%%%%%%%%%%%%%%%%%%
%%%%%%%%%%%%%%%%%%%%%%%%%%%%%%%%%%%%%%%%%%%%%%%%%%%%%%%%%%%%%%%%%

%%%%%%%%%%%%%%%%%%%%%%%%%%%%%%%%%%%%%%%%%%%%%%%%%%%%%%%%%%%%%%%%%
%%%%%%%%%%%%%%%%%%%%%%%%%%%%%%%%%%%%%%%%%%%%%%%%%%%%%%%%%%%%%%%%%

For WFC3, the \_raw images from the instrument were manually reprocessed with the latest calibration reference 
files that have been released as alpha and beta releases on the WFC3 web pages, often not yet ingested into the 
automated pipeline.  The primary steps involved in both the automated and manual 
processing of \_raw data from the CCDs included a bias subtraction, dark subtraction, flat fielding 
correction and gain conversion, as well as several other steps that populate the data quality and 
error arrays and calculate photometric header keywords.  The WFC3/IR array is not a CCD and requires 
additional steps such as the ``up-the-ramp'' fitting to image units of count-rate.  The manual processing 
of these new data with an updated version of the {\it calwfc3} pipeline allows us to intervene at any stage 
and apply the latest reference files.  Further details of the processing steps are provided in the WFC3 
Data Handbooks, \cite{rajan10}. 

The primary goal of this study is to find and measure the faintest possible stars in the imaging fields.  These stars 
often represent 0.2$\sigma$ events relative to the background in individual images, and can 
only be seen by carefully analyzing all the of images together.  There are two ways to do this.  One, described 
in Anderson \& King (2000) and Anderson et al.\ (2008a; 2008b) involves examining all of the images together to find the places 
in the field where a statistically significant number of images registered a marginal detection.  The pixels 
in all of the images in the vicinity of this location are then fit simultaneously to yield the most probable 
flux for the source, with careful attention to the PSF and S/N in each pixel (e.g., Lauer 1999a).  The final 
photometric measurements from this treatment of \_flt images has led to exquisite CMDs for the bulk of the Milky Way 
globular clusters (e.g., Anderson et~al.\ 2008a; 2008b), and the discovery of multiple stellar populations in these 
systems (e.g., Piotto et~al.\ 2007 and Piotto 2009 and references therein).

An alternate approach for finding these faintest sources involves combining the multiple dithered \_flt 
images for a given filter into a stacked image, and to perform photometric and astrometric measurements 
on the single coadded frame.  Several techniques exist for achieving the stacked image, from interlacing 
of input pixels from multiple individual images onto a finer grid of different output pixels (e.g., 
Lauer 1999b) to linearly shifting and adding pixels on to the subsampled grid.  Although interlacing can 
provide optimum combination of the images, the approach is limited in practice by several effects related 
to pointing accuracy and geometric distortion, as discussed in \cite{fruchter09}, \cite{koekemoer02},  
and references therein.  An improved method of image combination specifically for undersampled {\it HST} 
data is called MultiDrizzle \citep{fruchter02}.  In this algorithm, the individual pixels from the \_flt 
images (the input pixels) are rained down (or "drizzled") onto a common reference frame, which is often 
super-sampled relative to the input frames.  To do this, the location of each pixel is first corrected 
for geometric distortion and then the entire frame of pixels is shifted, rotated, and stretched so 
that each pixel is mapped into its proper place in the master frame.  The size of each input pixel is 
then shrunk to a smaller drop size than its original footprint (as described in Fruchter \& Sosey 2009).  
Although the MultiDrizzle approach still smears the final output image 
with the convolution of the drop size and the output pixels, in cases where the input images are 
well dithered, it is able to provide improved sampling of the PSF relative to the individual 
input images.  This is especially important on wide-field {\it HST} cameras such as ACS, WFC3/UVIS, 
and WFC3/IR where the size of a native pixel is comparable to the full width at half maximum (FWHM) 
of the PSF.

To demonstrate that the scientific results are independent of analysis methods, we will 
analyze these data using both approaches discussed above.  In this paper, we describe the 
latter analysis using MultiDrizzle which reduces the 707 images collected to a single stacked, 
supersampled image in each of the handful of instrument, filter, and exposure time combinations 
(e.g., one for deep frames and one for shallow).  In a future paper, Anderson et~al.\ (2012, in prep.), 
we will describe the complementary reduction which involves less convolution and more optimal 
weighting, but requires carefully solving for the spatially and temporally varying PSF on each 
input image.  

We next describe the specific methods used to stack the ACS, WFC3/UVIS, and WFC3/IR data.

\subsection{MultiDrizzling the ACS/WFC Primary Field} \label{MDACS}

%As discussed in the observational design in Section~\ref{acs} and summarized in Table~1, the ACS 
%primary field consists of 117 $F606W$ and 125 $F814W$ deep exposures of length 1000 -- 1500~s, 
%where five individual dithered observations are obtained in each filter in each visit.   
%Between the first 12 visits listed, the telescope was rolled by $\delta$ $\sim$ 20~degrees 
%so that the stars in the primary field would fall be sampled across different pixels and so that 
%the parallel field would map out an arc.  The remaining 12 visits were taken with an orientation that was held
%as constant as possible.  For the first set of rolled observations, the dithers were set to 
%pixel values of 3.75 and 7.5 arcsec, large enough to straddle the chip gap separating 
%the two CCDs on ACS.  For the second set of observations with small roll angles, both integer pixel and sub-pixel 
%dithers are combined.  The overall pattern therefore effectively randomized the location of sources in the 
%field of view, as illustrated in Figure~\ref{fig:context}.

The data processing was started by running a first pass of MultiDrizzle on the \_flt frames to generate 
individual output versions of all 117 $F606W$ and 125 $F814$ frames which have been corrected for 
the geometric distortion of the camera.\footnote{Although the STScI calibration pipelines also produced \_drz 
images, these are constructed with a default set of parameters for sky subtraction, rejection of bad 
pixels, and also do not include fine adjustments in image registration.  The overall quality of these 
combined images is much worse than what can be achieved by manually reprocessing the data.}  The 
distortion solution comes from the ACS calibration IDCTAB, u7n18502j\_idc.fits.  The 
accuracy of the WCS system and therefore the absolute alignment of these drizzled (``single\_sci'') 
images is limited by an intrinsic accuracy of $\sim$0.5~arcsec in the Guide Star Catalog.  To 
improve the fine alignment, a transformation is iteratively calculated between each of the 
distortion-corrected drizzled frames by isolating several hundred bright stars on each image and measuring 
their centroids using a Gaussian profile.  The transformation between these catalogs allows for linear 
offsets, rotations, and scale changes to refine the estimates of the Guide Star Catalog positions, 
account for the residual roll angle of the telescope from what was asked for, and factor in any 
breathing changes.  The solution was refined through successive matching of the common stars down to a 
tolerance of 0.3~pixels.  With the large number of stars available in the cluster, the final offsets 
provide an alignment of the individual images that is better than 0.01 pixel.

%%%%%%%%%%%%%%%%%%%%%%%%%%%%%%%%%%%%%%%%%%%%%%%%%%%%%%%%%%%%%%%%%
%%%%%%%%%%%%%%%%%%%%%%%%%%%%%%%%%%%%%%%%%%%%%%%%%%%%%%%%%%%%%%%%%

\begin{figure*}[ht]
\begin{center}
\leavevmode 
\includegraphics[width=12cm, angle=270]{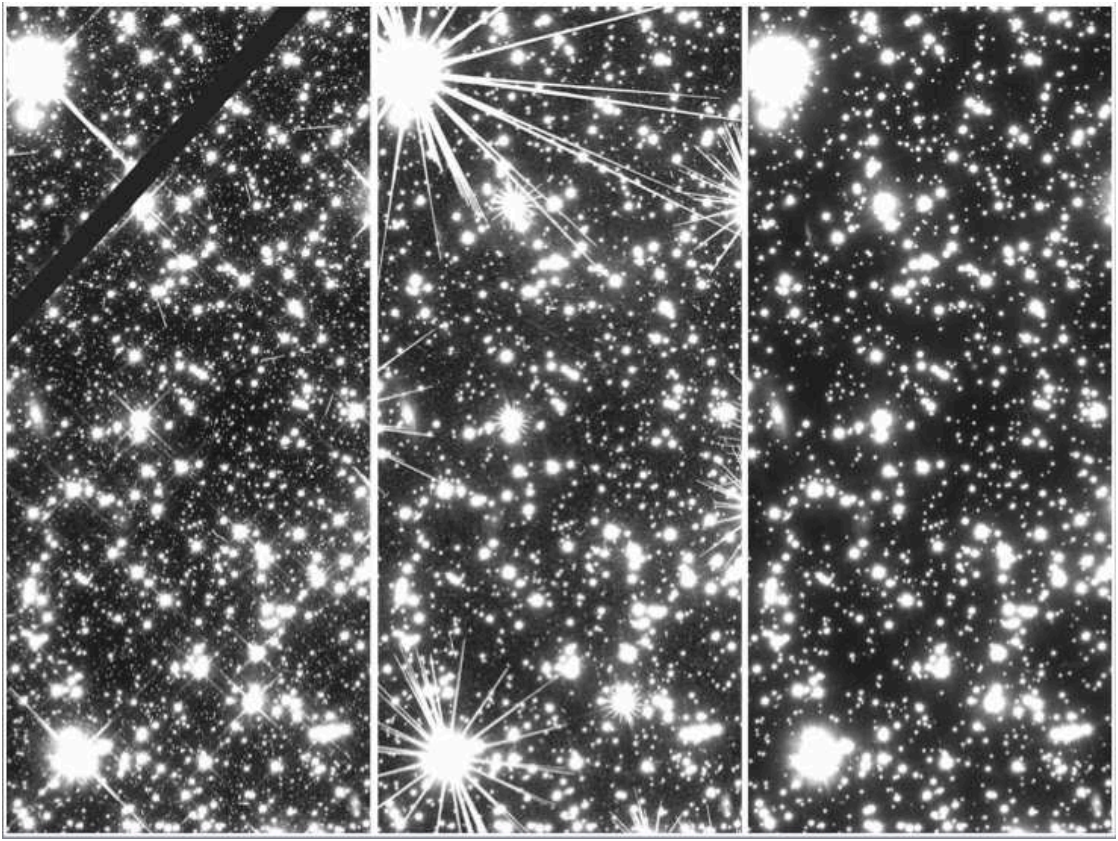}
\end{center}
\vspace{-0.3cm}
\caption{The quality of the final drizzled image for the ACS primary field is shown (right), 
relative to a single image that went into the stack (left) and a summed image of all 125 
input images (middle).  The single image exhibits many artifacts caused by bad pixels and cosmic rays, 
as well as the obvious chip gap separating the two CCDs on the ACS/WFC.  The most noticeable 
feature in the summed image is the large number of diffraction spikes surrounding the bright stars, 
each set corresponding to a unique roll angle of the observations.  Many of the cosmic rays were 
removed in generating this image to prevent it from being saturated with artifacts.  The final 
drizzled image in the right panel recovers information that is lost in the gap between the chips 
and is free of almost all artifacts.  The image shown here was generated with a Gaussian kernel, 
although we also experimented with kernels that gave a sharper final output image.  The one set of 
barely perceptible diffraction spikes is caused by the large number of exposures at approximately 
fixed roll angle for the Stare-field observations.  Each of the panels in this image extends about 
40 arcsec in the horizontal direction. \label{fig:ThreeImageMontage}}
\end{figure*}

%%%%%%%%%%%%%%%%%%%%%%%%%%%%%%%%%%%%%%%%%%%%%%%%%%%%%%%%%%%%%%%%%
%%%%%%%%%%%%%%%%%%%%%%%%%%%%%%%%%%%%%%%%%%%%%%%%%%%%%%%%%%%%%%%%%

These new offsets, rotations, and scales were supplied to MultiDrizzle in a second pass as a ``shift'' file 
and added to the WCS header information.  In this new stack of images, only real astrophysical sources are aligned, 
thereby providing a robust means to identify pixels that are affected by cosmic rays or are otherwise problematic 
(e.g., bad or hot pixels).  The different bits that had already been set in the \_flt images based on data 
quality arrays and bad pixel tables were largely ignored since the data provides a sensitive screening of 
those pixels that are specifically affecting the quality of our images.  For example, the multiple roll 
angles of the observations provide a means to largely eliminate diffraction spikes from bright stars, which 
march around their respective stars through the aligned sequence across different visits.  In addition to 
tweaking the location of each exposure with respect to the reference frame, offsets in the sky background 
of each exposure are accounted for and adjusted to match more closely that of the stack.  This provided 
a much better rejection of discordant pixel values.  Unlike extragalactic deep fields, the background of the 
images in this program is affected by the bright stars and so the purpose here 
is to ensure that any systematic offset level is accounted for prior to image combination.  The background on each 
image was calculated as the modal sky value after clipping pixels that are more than 5$\sigma$ deviating in 
multiple iterations.  A new bad-pixel mask was constructed by first producing a sigma-clipped median image in 
each of the two filters, and then reverse-drizzling (or ``blotting'') the median image back to the respective 
dither and roll angle of every original input image in the stack.  These blotted images were next compared to 
the input \_flt distorted images to statistically reject all bad pixels and cosmic rays.  This produces both 
clean versions of the input data as well as output masks with the bad pixels.

The final step of creating the image stack involved a third pass of MultiDrizzle to combine all of the 
input undistorted and aligned frames into a final image, utilizing the updated mask files to screen out 
bad pixels and cosmic rays.\footnote{We verified that drizzling yielded the expected gains in image depth 
by performing this step on smaller subsets of the data before executing the full stack.}  At this stage, 
we took further advantage of the multiple dither and roll 
angles to improve sampling of the PSF and to increase the spatial resolution of the final output image.  
For ACS, the FWHM of the PSF in the \_flt images is 1.7 -- 1.8 pixels in $F606W$ and $F814W$ (i.e., at the native 
resolution of 0.05 arcsec/pixel).  The pixels in all of the input images are shrunk and ``dropped'' down 
onto an output supersampled image with a scale of 0.03 arcsec/pixel.  For this process, we used a 
circular Gaussian kernel with ``pixfrac'' = 0.7 to distribute the flux onto the output grid.  This process 
conserves flux by computing the overlapping area between each of the input drops and output pixels, and 
dividing the flux that is drizzled down by appropriate weights.  The final combined image exhibits a 
well sampled PSF with $\sim$2.7~pixel FWHM.  The image spans $\sim$10,500 pixels in both directions.

In generating the final drizzled images, we also experimented with several different kernel functions for 
combination and choices of scale and pixfrac.  The Gaussian kernel leads to a slight ``blurring'' of the 
stars in the image relative to a square kernel (e.g., see Figure~\ref{fig:ThreeImageMontage}), but we 
found that our PSF modeling was slightly better in this case.  The balance between scale and pixfrac needs 
to ensure that the output image is not convolved with too large of an input-pixel drop size, yet that 
the drop size is large enough to provide output pixels that have data in them from a sufficient number 
of the input images (especially towards the edges of the frame).  The fidelity of the combined image 
was evaluated by examining a map of the relative weights of the output pixels.  If the pixels had 
been shrunk too much, large populations of output pixels would have little or no contribution 
from the input pixels, and this would lead to an increased standard deviation in the weight image.  If 
such large variation existed, photometric precision of astrophysical sources would be compromised.  According 
to \cite{fruchter09}, the ratio between the root-mean-square and median of this image (away from the 
edges) should be $\lesssim$20 -- 30\%.  This constraint is comfortably satisfied at all locations in the 
final ACS drizzled image given the large number of input images.  

A zoomed in view of a small portion of the final drizzled ACS field is presented in the right panel of 
Figure~\ref{fig:ThreeImageMontage}.  This image is for the $F814W$ filter and spans $\sim$40 arcsec 
in the horizontal direction.  The image is mostly free of artifacts and highlights a very 
clean view of the stellar populations, and also reveals a number of face-on and edge-on 
spiral and elliptical galaxies and some interacting pairs.  The one set of faint diffraction spikes 
around bright stars is caused by the approximately fixed orientation for half of the orbits (i.e., the 
roll angle for the Stare field).  For comparison, a single frame in the drizzled stack is shown in the 
left panel and shows many artifacts from bad pixels and cosmic rays, as well as the chip gap between the 
two CCDs.  A summed image of all of the aligned single frames is shown in the middle panel and shows 
diffraction spikes corresponding to each of the roll angles in the observations.  Many of the 
cosmic rays in this summed image were clipped for display purposes (i.e., to prevent it from being saturated 
with artifacts).

In the ACS/WFC deep exposures, stars with $\sim$19th magnitude were saturated on the detector.  The photometry and 
astrometry of these stars is therefore compromised.  To ensure that a complete stellar census is measured 
including bright giants in 47 Tuc, the observations that were obtained with shorter integration times 
were also reduced independently (discussed in Section~\ref{acs}).  The same methodology to that described above was 
adopted for these, but, given the more limited sampling information and resulting weight image statistics, 
the pixfrac was left at unity in the final combination.  This process produced a single stacked image in each of the 
two filters, for each of the 100~s, 10~s, and 1~s short exposures.

%deviation) which is less than 20\% of the median (midpoint) value. The 20\% weight 
%image threshold is a balance between the benefit of improving the image resolution 
%at the expense of increasing the noise in the background from resampling the pixels.
%In general, the ‘pixfrac’ should be slightly larger than the scale to allow some 
%spill over to adjacent pixels. 

%The output pixels in the final drizzled image are not independent of one another, 
%causing the noise in the output image to be correlated to some degree

\subsection{MultiDrizzling the WFC3/UVIS and IR \\ Parallel Fields} \label{MDWFC3}

The WFC3/IR array offers many advantages over conventional CCDs that work to our advantage, 
especially in cases where only a few images were obtained of a given field (i.e., all observations in 
this program except for the IR Stare field -- see Section~\ref{wfc3}).  Accumulated charge is electronically 
read out from the camera during an exposure through multiple non-destructive reads.  
Artifacts such as cosmic rays can therefore easily be identified and removed by fitting the count rate from 
these multiple reads (i.e., pixels affected by cosmic rays stood out in the read in which the hit impacted 
the array).    

All of the \_raw WFC3/UVIS and IR data were processed into calibrated \_flt frames using the latest 
reference files available on the WFC3 webpage (i.e., the new alpha-release flat fields).  The \_flt files 
were input into MultiDrizzle and processed using the same prescription described above for ACS (i.e., using the 
uab1537ci\_idc.fits IDCTAB).  This ensured an accurate alignment of all exposures in a given filter, 
sky subtraction, and the masking out of deviant pixels (using the ``minmed'' option instead of the median). 
The final image stack for WFC3/UVIS was generated 
at the native resolution of 0.04 arcsec/pixel, whereas the WFC3/IR images were slightly supersampled to a pixel 
scale of 0.09 arcsec/pixel to mitigate the severe undersampling of these data (native FWHM = 1.1 -- 1.2 pixels).  
A square kernel was used in the final image generation and the pixfrac was kept near unity, guided by an iterative 
process of performing photometry on the resulting stacked image and choosing the input set up that leads to the 
best defined CMDs (see below).

%%%%%%%%%%%%%%%%%%%%%%%%%%%%%%%%%%%%%%%%%%%%%%%%%%%%%%%%%%%%%%%%%
%%%%%%%%%%%%%%%%%%%%%%%%%%%%%%%%%%%%%%%%%%%%%%%%%%%%%%%%%%%%%%%%%

\begin{figure*}[ht]
\begin{center}
\leavevmode 
\includegraphics[width=13.0cm, angle=90]{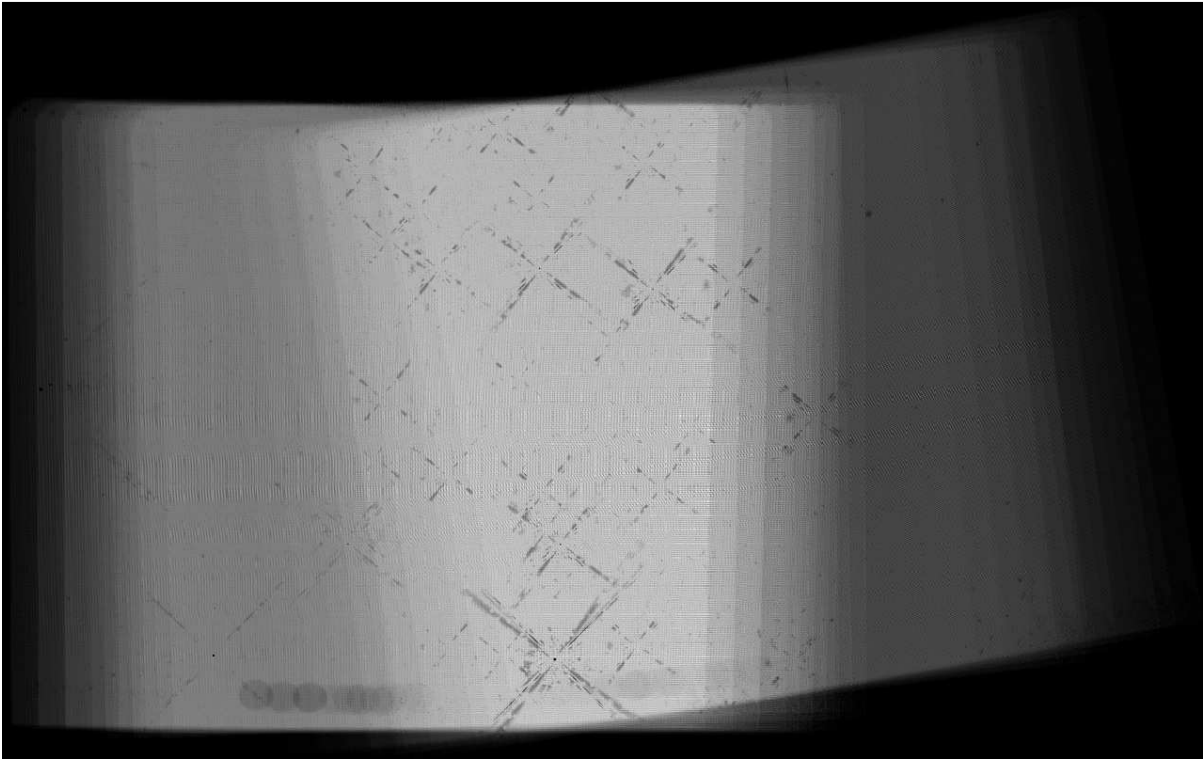}
\end{center}
\vspace{-1.5cm}
\caption{The weight image of the final combined WFC3/IR Stare field in $F160W$.  The shading 
indicates an effective exposure time map, where regions near the center of the image have 
80 overlapping independent images.  The ``slosh'' around the center is caused by small 
roll changes in the primary ACS observations over 59 orbits of the program.  These $\pm$5~degree 
roll changes result primarily from scheduling constraints and work to provide wider field 
coverage in the parallel observations.  Several artifacts can be seen in this image, such as 
diffraction spikes from bright stars (note, there is no bleeding of saturated pixels in the IR array) 
and the ``death star'' on WFC3/IR in the lower left -- a region of dead pixels.  As for the ACS 
primary field, this weight image was inspected to verify that the final choices of output pixel 
scale and drop size in MultiDrizzle were appropriate given the input dither pattern. \label{fig:weight}}
\end{figure*}

%%%%%%%%%%%%%%%%%%%%%%%%%%%%%%%%%%%%%%%%%%%%%%%%%%%%%%%%%%%%%%%%%
%%%%%%%%%%%%%%%%%%%%%%%%%%%%%%%%%%%%%%%%%%%%%%%%%%%%%%%%%%%%%%%%%

For the IR Stare field, the observations contained 38 $F110W$ and 80 $F160W$ deep exposures with many dithers 
and a small amount of roll.  The area on the sky mapped out by WFC3/IR during these 59 orbits is about 60\% 
larger than a single WFC3/IR field, as shown in an effective exposure time map (i.e., a ``weight'' image) 
in Figure~\ref{fig:weight}.  These 
data are more similar to the primary-field observations and can be combined to produce a higher-quality 
drizzled image than in the Swath.  The same techniques as described above were again followed to align and clean 
all of the input images using MultiDrizzle, and to combine them into the final stack.  The drizzled image was 
supersampled from the native resolution of 0.13 arcsec/pixel to 0.06 arcsec/pixel, with a Gaussian 
kernel with pixfrac = 0.70.  The FWHM on this image is $\sim$2.5~pixels.  As with the ACS primary field, 
the weight image of the Stare field was inspected to verify that the standard deviation in the output 
pixels was well below the median (see Figure~\ref{fig:weight}).  An example of the final drizzled image 
for both the Swath and Stare fields is shown in Figure~\ref{fig:montage}.

In addition to the deep observations, multiple short exposures were obtained on both WFC3 cameras to 
mitigate saturation (see Section~\ref{wfc3}).  For WFC3/UVIS, the short exposures consisted of two 
50~s observations in each filter.  These were drizzled together as described above.  For WFC3/IR 
the shorter exposures were analyzed, but they did not add any new information on top of what was recovered 
in the deep frames.  The reason for this is again related to the ``up-the-ramp'' fitting of the 
signal on the IR array through multiple non-destructive reads, the shortest of which have integration 
times of just a few seconds in which even the brightest stars do not saturate.  This is described more in 
Section~\ref{merging}.

%A key to understanding the use of pixfrac (Section 5.5.6) is to realize that a CCD 
%image can be thought of as the true image convolved ﬁrst by the optics, then by the 
%pixel response function (ideally a square the size of a pixel), and then sampled by a 
%delta-function at the center of each pixel. A CCD image is thus a set of point samples 
%of a continuous two-dimensional function. Hence the natural value of pixfrac is 0, 
%which corresponds to pure interlacing. Setting pixfrac to values greater than 0 causes 
%additional broadening of the output PSF by convolving the original PSF with pixels of 
%non-zero size. Thus, setting pixfrac to its maximum value of 1 is equivalent to 
%shift-and-add, the other extreme of linear combination, in which the output image PSF 
%has been smeared by a convolution with the full size of the original input pixels. 

\section{Photometry} \label{Photometry}

The image-processing steps described in the last section combined hundreds of individual 
exposures into a handful of drizzled images.  For the primary ACS field, we now 
have a single $F606W$ and $F814W$ well-sampled deep exposure and corresponding shallower 
images.  For the Swath and Stare fields, a montage of WFC3/UVIS (deep and 
shallow) and WFC3/IR (deep) exposures over 13 different fields have been produced and also 
drizzled together as shown in Figure~\ref{fig:montage}.  The observations in the IR Stare field 
are well sampled, but the Swath is undersampled.

To measure the photometry, astrometry, and morphology of all sources, the standalone 
versions of the DAOPHOT~II and ALLSTAR photometry programs were used \citep{stetson87,stetson94}. 
These programs offer flexibility in modeling the PSF with a range of analytic functions 
and account for some of its spatial variations across the frame \citep{stetson92}.  These 
variations are known to be significant for HST cameras such as ACS/WFC \citep{anderson08a}, 
and are currently being characterized for WFC3.  Such PSF-fitting methods can improve the signal-to-noise 
of measurements over aperture photometry and also provide a much better characterization 
of stars that are either located close to other stars and have contamination from the neighbor 
and/or blended stars that directly overlap one another. 

% Although not believed to be a large 
%concern for HST data, severely undersampled data can also pose significant problems since most 
%of the flux sits in the central pixel and therefore the centroid of the light distribution is 
%difficult to measure.  In these cases, sharper images can lead to a loss of photometric 
%stability (Lauer 1999a), thereby amplifying the need for sub-pixel dithers to recover spatial 
%resolution.

\subsection{PSF Photometry in the ACS/WFC Primary Field} \label{PhotometryACS}

To analyze the well-sampled ACS/WFC drizzled images in each filter, a first pass of DAOPHOT 
was performed to yield positions of all possible detections in both deep images. 
These were taken to be sources that are at least 2.5$\sigma$ above the local sky.  
Photometry of these candidates was estimated in each filter independently, using an 
aperture with $R$ = 2~pixels.  The images in each filter were treated independently 
at every stage of the analysis, and therefore there is a transformation that connects them.  
This transformation was derived by first selecting several hundred bright stars and 
cross-correlating the lists to define an $x$ and $y$ offset, rotation, and scale.  Based 
on these initial values, the transformation was refined by feeding the cross correlation 
the full list of detections in each image and iterating the resulting solution by successively 
decreasing the matching radius tolerance down to 1.0 pixel.  These two lists were combined 
and a first CMD was generated consisting of $>$70,000 measurements.

Next, 1000 PSF candidate stars were selected from this merged list in each filter.  These stars 
were required to be isolated (no neighbors within 19 pixels), not 
saturated, and to have a magnitude and color that placed them on the dominant stellar 
sequences in the CMD.  The PSF of each image was calculated through an iterative 
method.  Initially, all 1000 stars were used to build a spatially constant PSF, which 
was represented by the sum of an analytic function and an empirical look-up table representing the 
correction of the function to the observed brightness of the average profile of 
stars.  For the analytic function, the sum of a Gaussian and a Lorentzian with 
five free parameters was used, as described in \cite{stetson92}.  After fitting the analytic 
function to the observed stellar profiles of the PSF stars, the DAOPHOT-reported root-mean 
square residuals ($\chi^2$) of the brightnesses was inspected and PSF stars that showed scatter 
more than twice the average were eliminated from the list.  This entire process was repeated a 
second time with a PSF that was allowed to vary linearly with position in the frame, a 
third time allowing for quadratic variations, and a fourth time allowing for third 
order polynomial variations (i.e., cubic).  The number of initial PSF candidates that 
passed these successive iterations was 881 for $F606W$ and 874 for $F814W$, and the 
final $\chi^2$ is $\lesssim$0.02 indicating that the analytical function fit the 
observed profiles to within about 2\%.  This residual, a systematic difference, 
formed the look-up table of profile corrections.

The final step of the processing applied the PSF defined above to the catalog of 
all sources in each image.  For this, ALLSTAR was used to perform both PSF-fitted 
astrometry and photometry, and also to retain morphological information of sources 
through the SHARP diagnostic.  SHARP provides an estimate of the concentration of 
the source with respect to the PSF, which gives us a way to discriminate between stars 
and galaxies.

The same procedure described above was repeated on each of the drizzled shallow 
exposures in the ACS primary field, yielding independent catalogs of the brighter 
stars.  

%%%%%%%%%%%%%%%%%%%%%%%%%%%%%%%%%%%%%%%%%%%%%%%%%%%%%%%%%%%%%%%%%
%%%%%%%%%%%%%%%%%%%%%%%%%%%%%%%%%%%%%%%%%%%%%%%%%%%%%%%%%%%%%%%%%

\begin{figure*}[ht]
\begin{center}
\leavevmode 
\includegraphics[width=12.0cm, angle=270]{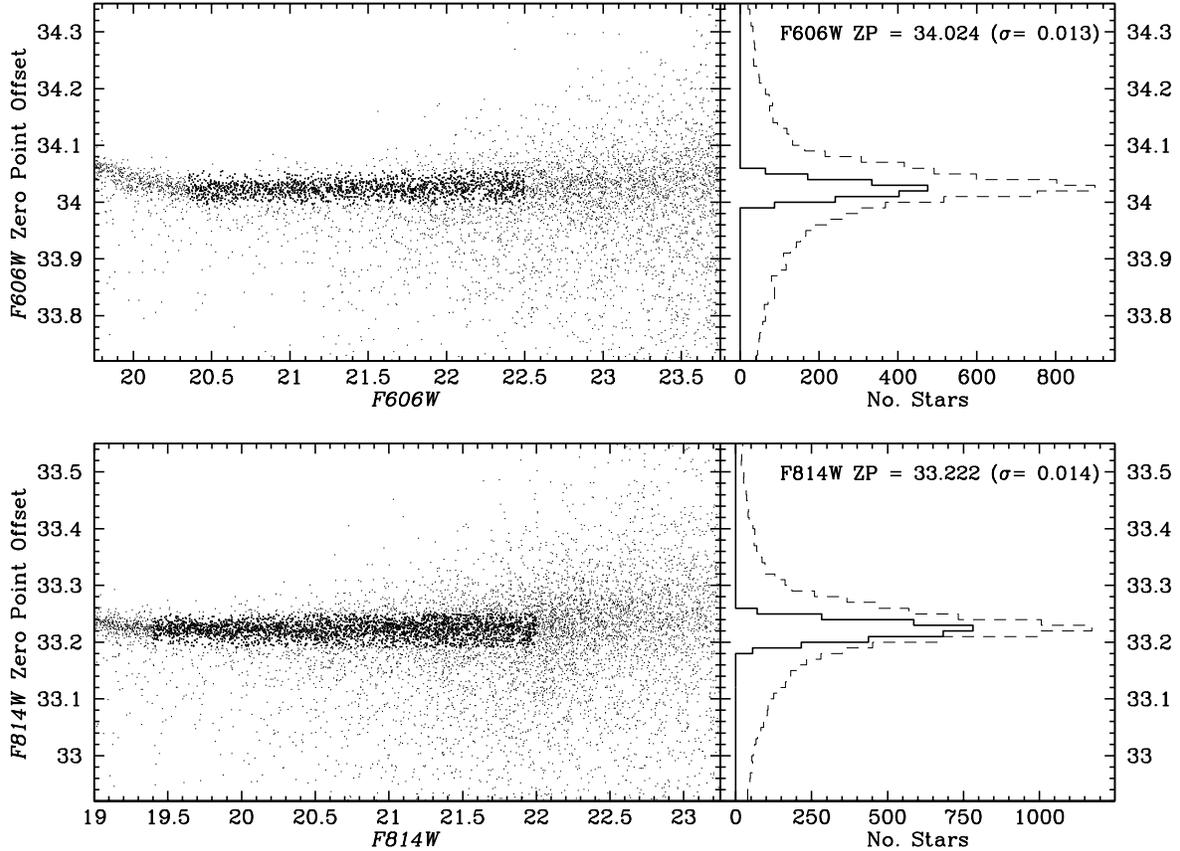}
\end{center}
\vspace{-0.7cm}
\caption{The derivation of the photometric zero points that are specific to the ACS primary field.  
The left panels show the difference between 5 pixel aperture photometry on a single drizzled image 
at the native scale and the PSF-fitted photometry on the super-sampled stacked image.  The 5 pixel 
photometry has been calibrated to the VegaMag photometric system using the ACS zero points (Bohlin~2007).  
Stars with poor measurements due to faint neighbors were eliminated (e.g., the bias towards lower values 
in the plot) and zero points were calculated from the resulting distribution (darker points).  These yield 
offsets of $F606W$ ZP = 34.024 and $F814W$ ZP = 33.222 for these reductions.  Histograms of the distributions 
are shown in the panels on the right.\label{fig:ZP}}
\end{figure*}

%%%%%%%%%%%%%%%%%%%%%%%%%%%%%%%%%%%%%%%%%%%%%%%%%%%%%%%%%%%%%%%%%
%%%%%%%%%%%%%%%%%%%%%%%%%%%%%%%%%%%%%%%%%%%%%%%%%%%%%%%%%%%%%%%%%

\subsection{PSF and Aperture Photometry in the WFC3 Parallel Fields} \label{PhotometryWFC3}

As discussed in Section~\ref{MDWFC3}, the WFC3 parallel Swath and Stare field observations 
have been reduced into a single drizzled image for each field and filter.  The focus of the 
Swath fields is to sample a wide area of the cluster periphery, and therefore there are very few 
overlapping images in any given filter (see Section~\ref{wfc3}).  The result is that we are unable 
to improve the sampling for these drizzled stacks.  As this investigation is one of the first 
studies of crowded regime photometry with WFC3, 
different photometric methods were tested on these images to see which yields the most 
accurate photometry.  This includes both aperture and PSF-fitted photometry on both 
the drizzled images and the \_flt images (which required a correction for the pixel area map -- 
Kalirai et~al.\ 2010b).

For the slightly undersampled WFC3/UVIS Swath observations, PSF-fitted photometry on coadded images 
with DAOPHOT~II and ALLSTAR produced sequences in the CMDs that extended fainter and were 
tighter than those produced with the other methods mentioned above.  The prescription used for this was 
identical to the discussion above on the ACS primary 
field.  The $\chi^2$ residuals from the analytic fit to the stellar profiles, after multiple 
iterations of the PSF, was measured to be 2 -- 4\% over the 13 fields.  A similar procedure was also 
performed independently on the shallower drizzled exposures.  

For WFC3/IR, the Swath data were too undersampled to accurately derive a PSF from the frame with 
DAOPHOT~II.  The photometry of all sources was therefore measured using an aperture 
with $R$ = 3.5~pixels, where each pixel is 0.09 arcsec in the resampled images.  
Our independent analysis of the \_flt images in the Swath fields for both cameras 
provide very similar photometric results to the drizzled images.

Unlike the WFC3/IR Swath fields, the Stare field was drizzled to an output scale of 0.06 
arcsec/pixel, and is therefore well sampled.  For this field, a spatially varying PSF with 
DAOPHOT~II was calculated as described above and was used with ALLSTAR to fit the stellar 
position, flux, and morphology of all sources.   

\subsection{Merging the Bright and Faint Starlists} \label{merging}

The bulk of the imaging exposures in this program were obtained with integration times of $>$1000~s, 
and were intended to optimize the signal-to-noise of faint stars in 47~Tuc.  The brightest cluster 
giants are more than 10 million times brighter than the faintest dwarfs, and therefore many stars 
saturated on both the ACS and WFC3/UVIS CCD observations.  Accurate photometry for these 
bright stars was measured from the short exposures described earlier.  The individual catalogs from each of the 
drizzled images with unique exposure times were ``stitched'' together by first finding well-measured, 
non-saturated, and isolated stars in each group and calculating a transformation between the 
coordinates of these stars.  The photometry of the bright frames was zeropointed to the deep frame 
using these common stars, and all stars in the bright frames which were not measured in the deep 
frame were mapped into the catalog after applying the transformation.  In this way the catalog 
contains uniform photometry and astrometry that is based on the deep drizzled image.

For the WFC3/IR frames, the multiple non-destructive reads provide accurate count-rates 
for most of the bright stars in the data set.  The deep observations were constructed with the 
STEP200 and STEP400 sample sequences with at least 12 ramps, and these included reads at 2.9~s, 5.9~s, 
8.8~s, ... 1399~s (see WFC3 Instrument Handbook -- Dressel et~al.\ 2010).  Although the effective read 
noise is larger for sources that have count rates estimated from just a few reads, this is overwhelmed 
by the photon statistics from such bright stars.  Therefore, with a single exposure time, we were 
able to measure stars spanning 13 magnitudes in the IR CMD.\footnote{The phase II program and APT 
design of these observations was submitted in December 2008, well before Servicing Mission 4 and 
any knowledge of instrument on-orbit photometric performance with limited numbers of reads.}

\subsection{Producing a Panchromatic Catalog} \label{panchromatic}

Thus far, the analysis has treated the WFC3 data from the UVIS and IR cameras separately.  
The WFC3/UVIS field of view is $\sim$60\% larger than the IR fields, and so the complete 
contiguous arc shown in Figure~\ref{fig:montage} (the IR fields) contains overlapping UVIS 
observations in $F390W$ and $F606W$.  A transformation was next calculated to map the drizzled 
IR astrometry onto the UVIS reference frame, accounting for small rotations, shifts, and the 
large plate-scale change of the two cameras (the native UVIS pixels are $>$2$\times$ smaller 
than even the supersampled IR pixels).  The result of this transformation is a catalog with common 
stars, measured in any number of multiple filter combinations.  A significant fraction of the 
stars have measurements in all four bandpasses, providing panchromatic imaging from 0.4 -- 
1.7~microns (see next Section).

\subsection{Photometric Zero Points} \label{zeropoints}

The photometric zero point of an instrument/filter combination is a convenient 
way to characterize the overall sensitivity.  Conventionally, the zero point represents 
the magnitude of a star-like object that produces one count per 
second within a given aperture (see Maiz~Apellaniz 2007).  For {\it HST} instruments, 
these zero points have been defined for each instrument by observing 
spectrophotometric standard stars with well measured fundamental parameters 
(e.g., $T_{\rm eff}$ and log($g$)) and STIS fluxes (Holtzman et~al.\ 1995; 
Sirianni et~al.\ 2005; Kalirai et~al.\ 2009a; 2009b).  The photometry of these isolated 
stars in large apertures is compared to the inferred sensitivity from the spectrum 
(or model) to provide the required calibration of the total system throughput of the 
filter (i.e., including the {\it HST} optics, detector quantum efficiency, filter 
transmission function, and many other components).

To place the 47~Tuc photometry on an absolute scale, the brightnesses of isolated stars on a 
single drizzled image are re-measured.  This (single\_sci) image has the geometric 
distortion removed but has not been otherwise altered (e.g., rescaled).  
Although the default photometric zero points are typically given for an ``infinite'' aperture that 
sums the total flux of the star, this can be rescaled to a smaller aperture using the 
enclosed energy curve for the respective camera (see Sirianni et~al.\ 2005 for ACS and 
Hartig~2009a; 2009b; Kalirai et~al.\ 2009a; 2009b for WFC3/UVIS and IR).\footnote{The 
flux in an infinite aperture is calculated by first measuring photometry of the isolated 
standard out to several arcseconds (e.g., $\sim$100~pixels on ACS/WFC and WFC3/UVIS) and then 
applying a small correction (typically $\lesssim$2\%) based on a model of the enclosed energy 
curve.}  This rescaling ensures that the photometry has sufficient signal-to-noise and is not 
compromised by neighboring stars in the cluster.  

An example of the zero-point derivation for the ACS/WFC primary field photometry is illustrated in 
Figure~\ref{fig:ZP}.  For this, the photometry of stars in a single drizzled image at the native scale 
in each filter was measured using an aperture with radius 5~pixels (0.25 arcsec), and the infinite 
ACS VegaMAG zeropoint of 26.4060 in $F606W$ and 25.5199 in $F814W$ was added.\footnote{The magnitude of a star with flux 
$f$ in the VegaMAG system is simply $-$2.5log($f$/$f_{\rm Vega}$), where $f_{\rm Vega}$ is the calibrated 
spectrum of the star Vega.}  Next, a correction of $-$0.151 and $-$0.171 was applied to translate the 
zero point to the aperture size based on the enclosed energy curve.  The difference between this 
calibrated photometry and the final PSF-fitted photometry from Section~\ref{PhotometryACS} is 
shown on the vertical axis as the zero point specific to our measurements.  The scatter in this 
diagram is driven by faint stars and cosmic rays that fall within the 5~pixel radius on 
the single image, so the distribution was $\sigma$ clipped to isolate the tight sequence of ``clean'' 
measurements (shown as larger points).  Saturated stars at the bright end of the distribution, with 
$F606W \lesssim$ 20 and $F814W \lesssim$ 19, were also eliminated.  The final zero point offsets, as 
measured from a few thousand stars, are 34.024 ($\sigma$ = 0.013) in $F606W$ and 33.222 ($\sigma$ = 
0.014) in $F814W$ and have errors $<$0.001 mag.

A similar analysis was performed on the WFC3/UVIS and IR Swath and Stare fields, using the photometric 
zero points published in \cite{kalirai09a} and \cite{kalirai09b} for an aperture of radius 10~pixels 
(UVIS) and 3~pixels (IR).  The derived zero points for these reductions are summarized in Table~4.
%For WFC3/UVIS, we measure $F390W$ ZP = 32.851 ($\sigma$ = 0.028) and $F606W$ 
%ZP = 33.697 ($\sigma$ = 0.023).  For the WFC3/IR Swath fields, we measure $F110W$ ZP = 33.604 
%($\sigma$ = 0.013) and $F160W$ ZP = 32.197 ($\sigma$ = 0.011) and for the Stare field we measure 
%$F110W$ ZP = 33.673 ($\sigma$ = 0.015) and $F160W$ ZP = 32.253 ($\sigma$ = 0.016).

%%%%%%%%%%%%%%%%%%%%%%%%%%%%%%%%%%%%%%%%%%%%%%%%%%%%%%%%%%%%%%%%%%%%%%%%%%%%%%%%%%%%
%%%%%%%%%%%%%%%%%%%%%%%%%%%%%%%%%%%%%%%%%%%%%%%%%%%%%%%%%%%%%%%%%%%%%%%%%%%%%%%%%%%%
%%%%%%%%%%%%%%%%%%%%%%%%%%%%%%%%%%%%%%%%%%%%%%%%%%%%%%%%%%%%%%%%%%%%%%%%%%%%%%%%%%%%

\begin{table}
\begin{center}
\caption{Photometric Zero Points for Final Reductions}
%\vskip 0.3cm
\begin{tabular}{lcc}
\hline
\hline
\multicolumn{1}{c}{Filter} & \multicolumn{1}{c}{Zero Point} & \multicolumn{1}{c}{$\sigma$} \\
\hline
{\bf ACS} \\
$F606W$ & 34.024 & 0.013 \\
$F814W$ & 33.222 & 0.014 \\
\\
{\bf WFC3/UVIS} \\
$F390W$ & 32.851 & 0.028 \\
$F606W$ & 33.697 & 0.023 \\
\\
{\bf WFC3/IR Swath Fields} \\
$F110W$ & 33.604 & 0.013 \\
$F160W$ & 32.197 & 0.011 \\
\\
{\bf WFC3/IR Stare Fields} \\
$F110W$ & 33.673 & 0.015 \\
$F160W$ & 32.253 & 0.016 \\
\hline
\end{tabular}
\label{table2}
\end{center}
\end{table}

%%%%%%%%%%%%%%%%%%%%%%%%%%%%%%%%%%%%%%%%%%%%%%%%%%%%%%%%%%%%%%%%%%%%%%%%%%%%%%%%%%%%
%%%%%%%%%%%%%%%%%%%%%%%%%%%%%%%%%%%%%%%%%%%%%%%%%%%%%%%%%%%%%%%%%%%%%%%%%%%%%%%%%%%%
%%%%%%%%%%%%%%%%%%%%%%%%%%%%%%%%%%%%%%%%%%%%%%%%%%%%%%%%%%%%%%%%%%%%%%%%%%%%%%%%%%%%

%CHI, is the ratio of the observed pixel-to-pixel scatter in the fitting residuals 
%to the expected scatter, based on the values of readout noise and the photons per 
%ADU which you specified in your aperture photometry table.  If your values for the 
%readout noise and the photons per ADU are correct, then in a plot of CHI against 
%derived magnitude (e.g. Stetson and Harris 1988, A.J. 96, 909, Fig. 28), most stars 
%should scatter around unity, with little or no trend in CHI with magnitude (except 
%at the very bright end, where saturation effects may begin to set in.

%SHARP is an image-peculiarity statistic, vaguely related to the intrinsic (i.e. outside
%the atmosphere) angular size of the astronomical object: effectively, SHARP is a 
%zero-th order estimate of the square of the quantity (actual one-sigma 
%half-characteristic-width of the astronomical object as it would be measured outside 
%the atmosphere, in pixels):  SHARP2 ~ sigma^2 (observed) - sigma^2 (PSF).
%Objects with SHARP significantly greater than zero are probably galaxies or 
%unrecognized doubles; objects with SHARP significantly less than zero are probably 
%bad pixels or cosmic rays.

%%%%%%%%%%%%%%%%%%%%%%%%%%%%%%%%%%%%%%%%%%%%%%%%%%%%%%%%%%%%%%%%%
%%%%%%%%%%%%%%%%%%%%%%%%%%%%%%%%%%%%%%%%%%%%%%%%%%%%%%%%%%%%%%%%%

\begin{figure*}[ht]
\begin{center}
\leavevmode 
\includegraphics[width=12.0cm, angle=270]{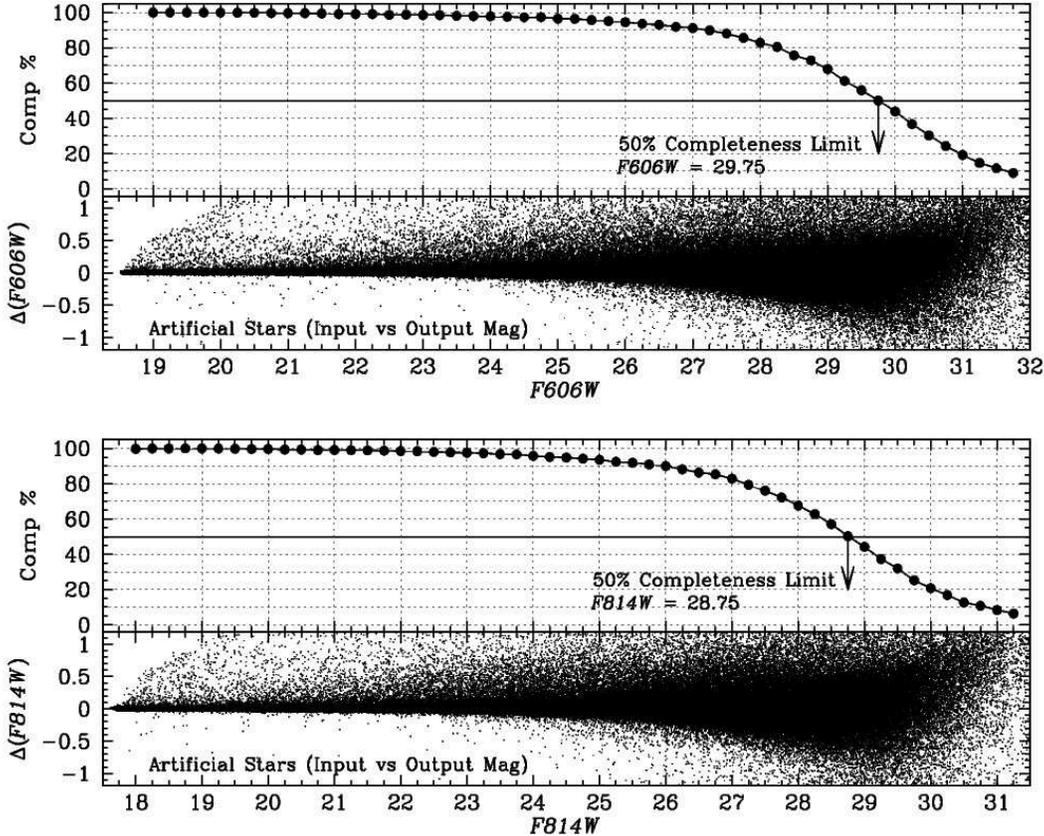}
\end{center}
\vspace{-0.5cm}
\caption{The completeness fraction of the ACS/WFC stacked images is measured through artificial-star tests.  
The top panels illustrate the completeness curve and photometric error distribution (e.g., 
input artificial star magnitude vs recovered magnitude) for the $F606W$ image and the bottom panels 
illustrate the same for the $F814W$ image.  The 50\% completeness limits are measured to be $F606W$  = 29.75 
and $F814W$ = 28.75.  Similar tests were also performed on each field of the Swath and Stare, in both the 
WFC3/UVIS and WFC3/IR data.\label{fig:ART}}
\end{figure*}

%%%%%%%%%%%%%%%%%%%%%%%%%%%%%%%%%%%%%%%%%%%%%%%%%%%%%%%%%%%%%%%%%
%%%%%%%%%%%%%%%%%%%%%%%%%%%%%%%%%%%%%%%%%%%%%%%%%%%%%%%%%%%%%%%%%

\section{Artificial-Star Tests}

We characterize the photometric and astrometric error distribution of the final photometry in our images, as well as the 
completeness of the data reductions by performing an extensive set of artificial-star tests.  These tests are 
tailored to each of the individual data products, including all six filters observed with ACS, WFC3/UVIS, and WFC3/IR, 
and are computed separately for each individual field to fully account for crowding variations due to the wide-field 
coverage of 47~Tuc.  

The artificial stars are modeled from the PSF of each individual field, and scaled to reproduce the complete luminosity 
range of real stars in the deep drizzled stacks of each filter.  The fraction of stars injected into each image was set to 
$\lesssim$5\% of the total number of stars in the image, so as to not induce incompleteness due to crowding in the 
tests themselves. 500 trials were generated for each test combination (e.g., the observations in WFC3/UVIS field Swath~4, 
in the $F390W$ filter) to form the input grid of artificial starlists and resulting images.  These images were next subjected 
to the photometric routines that were applied to the actual drizzled images, using identical criteria.  The stars were 
recovered blindly and automatically cross-matched to the input starlists containing actual positions and fluxes.

The artificial-star tests from this program can be used to investigate a number of questions regarding the fidelity of 
our reductions.  For example, these tests are essential to the derivation of the stellar mass function of 47~Tuc.  
We highlight one aspect of the tests in Figure~\ref{fig:ART} by presenting the incompleteness curve for our ACS primary 
field $F606W$ and $F814W$ observations, as well as the photometric error distribution (e.g., artificial star input 
vs output magnitude).  These results 
indicate that the 50\% completeness fraction of the ACS data is $F606W$ = 29.75 and $F814W$ = 28.75.  The completeness 
limits in the WFC3/UVIS and IR data vary by more than one magnitude depending on the field observed.  Specific details will 
be presented along with the corresponding scientific investigations in future papers.  As a reference point, in the Swath 3 
field, the the 50\% completeness is $F390W$ = 28.3, $F606W$ = 28.8, $F110W$ = 26.2, and $F160W$ = 25.1.

The total number of artificial stars generated in these tests exceeds 2 million and total 5 TB of disk space. 

\section{Color Magnitude Diagrams} \label{CMDs}

\begin{figure*}[ht]
\begin{center}
\leavevmode 
\includegraphics[width=12.0cm, angle=270]{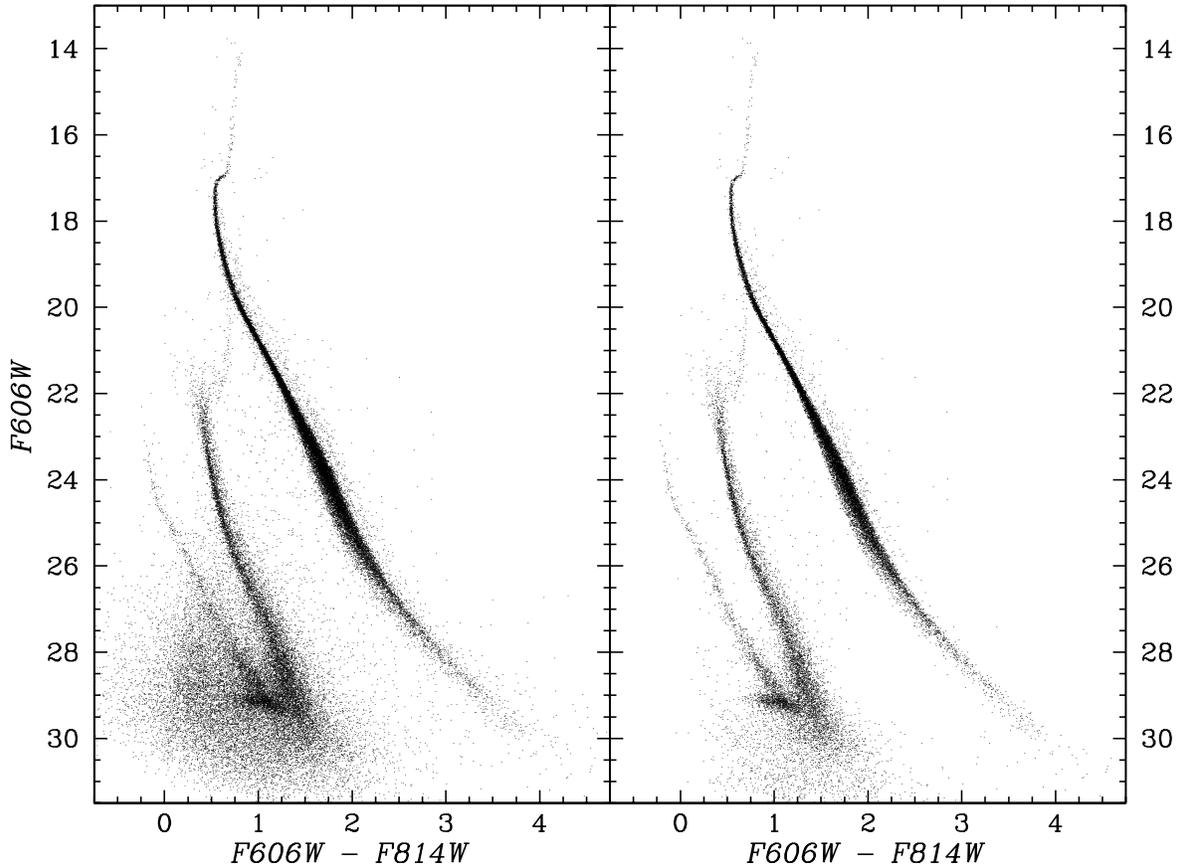}
\end{center}
\vspace{-0.5cm}
\caption{The CMD of all stars along the sightline in the ACS primary field, constructed from 
PSF-fitted photometry of a single deep supersampled drizzled image in each of the $F606W$ and $F814W$ 
filters (plus corresponding coadded shorter exposures).  The left panel 
illustrates all objects that were detected in both filters and the right panel shows those objects 
that pass a mild photometric error cut.  The CMD reveals a very well defined 47~Tuc main-sequence 
extending from bright giants at $F606W$ = 14 down to faint, low-mass hydrogen burning dwarfs 
at $F606W$ = 30, as well as a rich white-dwarf cooling sequence extending over 6 magnitudes in 
the faint-blue part of the diagram with a sharp hook to the blue at $F606W$ $\sim$ 29.  The deep 
imaging also penetrates past 47~Tuc to reveal over 10,000 stars in the background SMC dwarf galaxy.  The  
SMC population behind the cluster exhibits a very well defined red-giant branch, multiple 
main-sequence turnoffs, and a rich main-sequence extending down to $>$30th magnitude 
($M \lesssim$ 0.2~$M_\odot$).\label{fig:ACSCMD}}
\end{figure*}

CMDs from the PSF-fitted photometry of the final drizzled images in the UV, 
visible, and IR are presented in Figures~\ref{fig:ACSCMD} and \ref{fig:WFC3CMD}, and a panchromatic 
CMD over the full wavelength from 0.4 -- 1.7 microns in Figure \ref{fig:panchCMD}.  
These CMDs provide a convenient summary of the observations in our 121 orbit program, 
and the quality of the data and data analysis.  The primary features and depth of the CMDs is discussed 
here in relation to the program's science goals, which will be explored in greater detail in 
future papers.

%%%%%%%%%%%%%%%%%%%%%%%%%%%%%%%%%%%%%%%%%%%%%%%%%%%%%%%%%%%%%%%%%
%%%%%%%%%%%%%%%%%%%%%%%%%%%%%%%%%%%%%%%%%%%%%%%%%%%%%%%%%%%%%%%%%

%%%%%%%%%%%%%%%%%%%%%%%%%%%%%%%%%%%%%%%%%%%%%%%%%%%%%%%%%%%%%%%%%
%%%%%%%%%%%%%%%%%%%%%%%%%%%%%%%%%%%%%%%%%%%%%%%%%%%%%%%%%%%%%%%%%

\subsection{ACS Primary Field Visible Light Color Magnitude Diagram}

The CMD shown in Figure~\ref{fig:ACSCMD} includes all stars detected in both filters 
of the ACS primary field, and reveals three clear stellar populations.  The diagram on the 
right shows a subset of the data that pass a mild photometric error cut to eliminate poor 
measurements.  The rich sequence on the right of this diagram represents the 47~Tuc red 
giant branch, turnoff, and main-sequence, extending from $F606W$ = 14 -- 30.  This sequence 
includes 47~Tuc stars ranging from bright giants, through 
the main-sequence turnoff, down to low mass hydrogen burning dwarfs with $M$ $<$ 
0.10~$M_\odot$.  The main sequence exhibits variations in its thickness, an obvious 
binary sequence except at the faintest magnitudes, and mild evidence for a ``trickle'' 
of bluer stars on the lower main-sequence at $F606W >$ 26.  The interpretation of these 
features will follow from a full simulation of the main-sequence using updated stellar 
evolution models by our team.

The second parallel sequence in this CMD below the 47~Tuc main-sequence represents resolved 
giant and dwarf stars in the background SMC dwarf galaxy.  The photometry of this nearby 
galaxy extends down to $F606W$ $>$ 30 and the catalog contains over 10,000 stars.  These data 
will enable studies of the mass function of this galaxy's outer population 
down to stars with $M \sim$ 0.15 -- 0.20~$M_\odot$.  The main-sequence turnoff also indicates 
evidence of multiple turnoffs, confirming an extended star formation history in this remote 
field.  The comparison of the deep mass function of the SMC and the field halo population of 
the Milky Way will reveal important insights on the formation history of low mass stars with 
very different metallicities.

The primary science goal of this project is to measure the white-dwarf cooling age of 47~Tuc, 
and compare it to our recent measurements of M4 and NGC~6397 (Hansen et~al.\ 2004; 2007).  
Bluer than the SMC sequence, the ACS CMD reveals a rich white-dwarf population of the cluster.  
Towards the bottom of the cooling sequence near $F606W$ $\sim$ 29 and $F606W - F814W$ $\sim$ 1, 
a well defined ``hook'' in the remnant sequence is seen and is brighter than the faintest dwarfs 
of the SMC.  The morphological diagnostics in the reductions indicate that the bulk of the fainter, 
scattered population of objects in the CMD below the 47~Tuc white dwarfs and SMC main-sequence 
are distant galaxies.  These can be easily removed when analyzing the foreground populations.
Further details and investigation of this CMD and theoretical modeling efforts will be provided in 
Richer et~al.\ (2012, in prep.) and Hansen et~al.\ (2012, in prep.).  This work is expected to 
yield the most accurate age for 47~Tuc to date (e.g., by leveraging both white dwarfs and the 
main-sequence turnoff), the deepest mass function of the cluster, and new insights on cluster 
binarity and dynamics.

\subsection{WFC3 Parallel Field Color Magnitude Diagrams} \label{sec:CMD}

With the exception of the IR Stare field (59 orbits), all of the observations obtained with the 
WFC3 instrument in this program had integration times of one orbit in the $F390W$, $F606W$, 
and $F110W$ filters (two exposures each), and two orbits in the $F160W$ filter (four exposures).  
Most of the fields targetted relatively uncrowded regions of 47~Tuc, as shown in 
Figure~\ref{fig:montage}.  The observations therefore span a unique region of parameter space, 
not being limited by crowding yet sampling tens of thousands of cluster and 
background stars over the $>$60 arcmin$^2$ area in the 13 contiguous fields.

\subsubsection{WFC3/UVIS CMD} \label{sec:wfc3uvisCMD}

\begin{figure*}[ht]
\begin{center}
\leavevmode 
\includegraphics[width=12.0cm,angle=270]{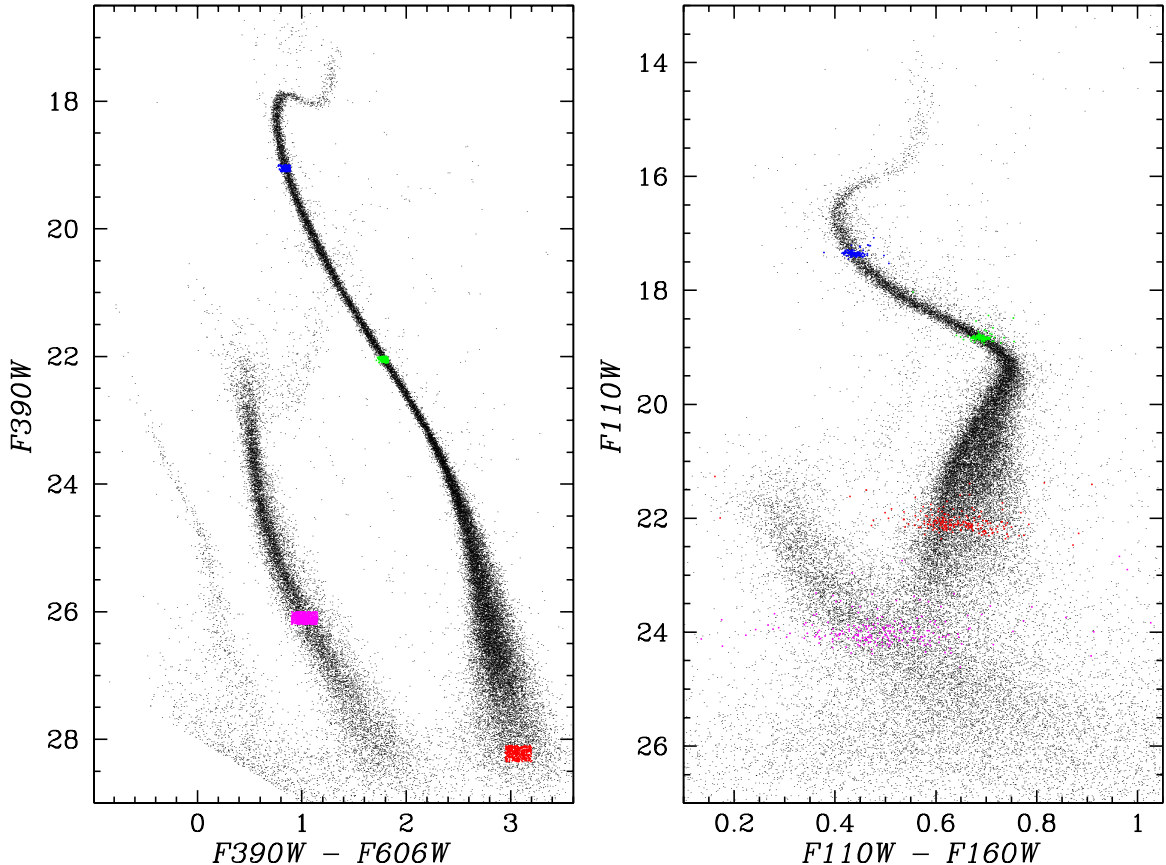}
\end{center}
\vspace{-0.5cm}
\caption{CMDs based on the Swath field observations with WFC3/UVIS (left) and WFC3/IR (right) exhibit 
very distinct morphologies and impressive depths considering the exposure time was a single orbit in 
each of the $F390W$, $F606W$, and $F110W$ filters, and two orbits in the $F160W$ filter (i.e., observations 
in the 59 orbit IR Stare Field are excluded here).  The UVIS CMD 
clearly highlights the same three populations of stars shown in the ACS primary field (Figure~\ref{fig:ACSCMD}); 
a rich sequence of 47~Tuc red giant branch and main-sequence stars, a background SMC dwarf main-sequence, 
and a white dwarf cooling sequence stretching over 7 magnitudes.  The data extend to beyond 28th magnitude 
in the $F390W$ filter.  The IR CMD, over a narrow color baseline of $<$1~magnitude, shows the rich 47~Tuc 
main-sequence extending to fainter and redder colors down to $F110W$ = 19.2, beyond which it sharply 
``kinks'' back to the blue and extends another 4 magnitudes to very low masses.  
Below the 47~Tuc population, the rich main-sequence of the SMC emerges with the opposite slope.  
To aid in comparing the morphology of the CMDs, the same stars in both panels are color coded along 
three locations of the 47~Tuc main-sequence and one location of the SMC main-sequence.
Further discussion of these CMDs, which combine observations over 13 contiguous fields and 
contain $\sim$50,000 stars, is provided in Sections \ref{sec:wfc3uvisCMD} and 
\ref{sec:wfc3irCMD}.\label{fig:WFC3CMD}}
\end{figure*}

The UVIS CMD of the Swath mosaiced fields is shown in the left panel of Figure~\ref{fig:WFC3CMD}, 
and contains $\sim$50,000 stars.  The CMD presents a very clean characterization of the 47~Tuc 
and SMC main-sequences down to $F390W$ = 28.5.  Whereas the 47~Tuc main-sequence is extremely 
sharp at $F390W$ = 22 -- 23, the SMC turnoff exhibits evidence of multiple splittings in this 
remote field indicative of an extended star formation history.  The WFC3/UVIS data also 
uncover a very rich white-dwarf cooling sequence with over 500 stars, extending almost 7 
magnitudes from $F390W$ = 21.2 (three stars) down to 28th magnitude.  

This rich population of {\it bright} white dwarfs in a co-eval stellar population represents a 
valuable sample for testing physical conditions and processes in degenerate objects.  The cooling timescale of 
young white dwarfs is strongly dependent on the rate at which neutrinos are radiated out of the 
degenerate core, and two of the primary factors that control this radiation are the density and 
temperature of the star \citep{itoh96}.  The observed luminosity functions will test the 
cooling physics of these stars, and even more so when coupled to independent constraints on 
the temperature and gravity of the stars from ground based spectroscopy.  These outer fields 
of 47~Tuc are relatively sparse, and the white dwarfs are brighter than those observed with 
Keck/LRIS in Messier 4, observations that successfully led to the first direct constraints 
on the mass distribution of population II white dwarfs \citep{kalirai09c}.  It is worth 
stressing that the exposure time of these observations was $\sim$2700~seconds in each filter, 
split over one dither position (i.e., two images per pointing).

%%%%%%%%%%%%%%%%%%%%%%%%%%%%%%%%%%%%%%%%%%%%%%%%%%%%%%%%%%%%%%%%%
%%%%%%%%%%%%%%%%%%%%%%%%%%%%%%%%%%%%%%%%%%%%%%%%%%%%%%%%%%%%%%%%%

%%%%%%%%%%%%%%%%%%%%%%%%%%%%%%%%%%%%%%%%%%%%%%%%%%%%%%%%%%%%%%%%%
%%%%%%%%%%%%%%%%%%%%%%%%%%%%%%%%%%%%%%%%%%%%%%%%%%%%%%%%%%%%%%%%%

\subsubsection{WFC3/IR CMD} \label{sec:wfc3irCMD}

A review of the astronomical literature shows that few space-based studies of nearby resolved stellar 
populations have used the IR bandpasses to test stellar evolution and structure theory for low mass 
stars.  For {\it HST}, the previous generation IR imaging camera NICMOS was not heavily used to image 
populations such as globular clusters on account of its small field of view.  For {\it Spitzer}, the Infrared 
Array Camera (IRAC) has both sensitivity over the 3 -- 9.5 micron wavelength range and a wide field of 
view, but with much lower throughput and 10$\times$ larger pixels than {\it HST} (resulting in poor 
performance in crowded regimes).  The IR channel of WFC3 therefore fills an important gap in our current observational 
potential, by providing 1) a large field of view of several square arcmin, 2) excellent sensitivity 
(Kalirai et~al.\ 2009b), 3) a relatively small pixel size leading to high resolution sampling (0.13~arcsec), 
and 4) excellent temporal and spatial photometric stability ($>$99\% -- Kalirai et~al.\ 2011).\footnote{The 
spatial stability and new on-orbit flat fields are described at 
{\tt http://www.stsci.edu/hst/wfc3/analysis/ir\_flats}}

The right hand panel of Figure~\ref{fig:WFC3CMD} presents the WFC3/IR CMD of these same Swath fields 
(i.e., excluding the deep 59 orbit IR Stare field).  The CMD has a unique morphology with a 
strong kink at $F110W$ = 19.2, about 3 magnitudes below the main-sequence turnoff, and a 
continuation of the sequence below the kink by another 4 magnitudes where it becomes {\it bluer} 
with increasing faintness. This very sharp feature has been detected in two recent 
ground-based studies, though not with the clarity and statistical significance of these measurements 
\citep{sarajedini09,bono10}.  The kink feature is caused by collisionally-induced absorption by molecular 
hydrogen in the atmospheres of cool stars with $T_{\rm eff} \sim 4500$~K and was anticipated in the low 
mass stellar models of \citet[][see their Section 4.2 and Figure 7]{baraffe97}.  This feature is also seen on
the UVIS CMD as a change in slope of the lower main-sequence near $F390W$ = 24; the effect is clearly much 
stronger in the IR (note the color baseline) because the molecular absorption bands most dramatically influence 
near-IR wavelengths \citep{saumon94}.  Because the kink is largely insensitive to age and metallicity, it provides 
greater constraints on model fits than the optical CMD (Dotter et~al.\ 2012, in prep.).  It grants improved 
leverage on the age {\it and} a critical test of stellar models \citep[see, e.g., Section 5 of ][]{sarajedini09}.  
The morphology and location of the main-sequence below the kink also represents an important calibration to 
establish the masses of low-mass hydrogen burning stars and brown dwarfs in the Galactic field, where 
metallicity information is not available.  Finally, these IR observations of the complete stellar populations 
of 47~Tuc are an important data point for red population synthesis models at high metallicity.

Fainter than the kink, the main-sequence of 47~Tuc extends to $F110W$ $\gtrsim$ 23, below 
which a second sequence emerges with the opposite slope.  This represents the SMC main-sequence and it extends 
in our data set for $\sim$4 magnitudes in the IR filters.  The red-giant branch of the SMC can be seen extending 
upwards from the tip of the turnoff, through the 47 Tuc main sequence and up towards the bright-red end of the CMD. 

To demonstrate the power of studying globular clusters in the IR bandpasses with WFC3, we select a small 
number of stars on the lower main sequence of the UVIS CMD in the left panel of Figure~\ref{fig:WFC3CMD} 
(shown as red points with $F390W$ = 28.1 -- 28.3), and plot the same stars on the IR CMD in the right panel.
These cool stars, have colors of $F390W - F606W$ $>$ 3, and are therefore much brighter in the IR 
bandpasses.  The same stars map to the main-sequence at $F110W$ = 22.2, well above the faintest low mass 
stars that are detected and $\sim$4 magnitudes above the faintest SMC dwarfs.  As most clusters 
do not have such a rich population of background stars, WFC3/IR imaging with just a few {\it HST} orbits 
can characterize the complete color-magnitude relation of stars and map the main-sequence mass function 
down to the hydrogen burning limit.

\subsubsection{Panchromatic CMD from 0.4 -- 1.7 microns} \label{sec:panCMD}

\begin{figure*}[ht]
\begin{center}
\leavevmode 
\includegraphics[width=12.0cm, angle=270]{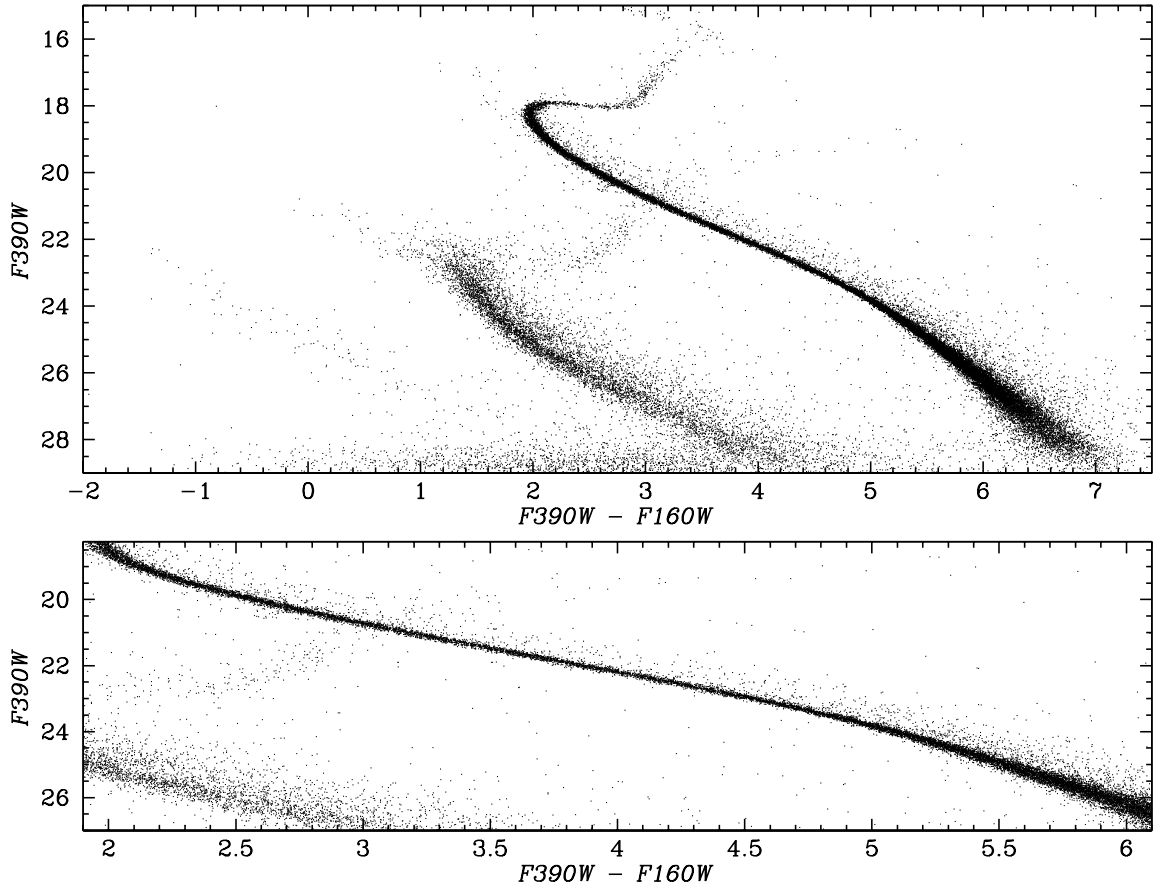}
\end{center}
\vspace{-0.5cm}
\caption{The panchromatic nature of this study is highlighted by constructing a CMD of the 
stellar populations over the widest baseline of $F390W - F160W$ (i.e., 0.4 -- 1.7 microns).
The combined WFC3/UVIS and IR data stretch the stellar populations over a color range of 
$>$9 magnitudes (top panel).  Despite their faintness in the IR, over 150 white dwarfs form a 
cooling sequence on this CMD.  The bottom panel focuses on the main-sequence of 47~Tuc, which is 
stretched over $>$4 magnitudes of color.\label{fig:panchCMD}}
\end{figure*}

The final CMD that is presented in Figure~\ref{fig:panchCMD} combines WFC3/UVIS and IR imaging over all 
Swath and Stare fields.  All objects that were well measured in both the bluest ($F390W$ 
on WFC3/UVIS) and reddest ($F160W$ on WFC3/IR) filters are selected for this.  The full CMD in the top panel extends 
over a color baseline of $>$9 magnitudes, from $F390W - F606W$ = $-$2 to 7.  A white-dwarf cooling sequence is 
mapped over this full baseline, enabling studies of the spectral energy distributions of the remnants, and 
investigations of objects that show IR excesses from accretion disks or stellar/sub-stellar companions (Woodley et~al.\ 
2012, submitted).  The 47 Tuc and SMC main-sequences are also very well populated in this CMD, 
which provides increased leverage to detect multiple splittings or turnoffs.  The bottom panel illustrates a 
closer view of the main-sequence of the cluster stretched over a color baseline of $>$4 magnitudes.  Unlike 
the case of NGC~2808 (e.g., D'Antona et~al.\ 2005; Piotto et~al.\ 2007; see Appendix), this CMD 
shows no signs of multiple populations for the combined photometric catalog.  One caveat to this statement 
is that, despite the broad wavelength range of this CMD, the $F390W$ filter sits at the red end of the UV 
spectrum and is therefore not sensitive to some elemental variations in the far UV.  In addition to 
the lack of clear splittings in the 47~Tuc CMD, the data indicate no strong spatially dependent variations 
of the thickness of the cluster sequence across fields separated by large radial distances in the Swath.  
These initial hints will be fully explored by our team in the near future through a complete study of the 
stellar and dynamical state of 47~Tuc from these data.

%%%%%%%%%%%%%%%%%%%%%%%%%%%%%%%%%%%%%%%%%%%%%%%%%%%%%%%%%%%%%%%%%
%%%%%%%%%%%%%%%%%%%%%%%%%%%%%%%%%%%%%%%%%%%%%%%%%%%%%%%%%%%%%%%%%

%%%%%%%%%%%%%%%%%%%%%%%%%%%%%%%%%%%%%%%%%%%%%%%%%%%%%%%%%%%%%%%%%
%%%%%%%%%%%%%%%%%%%%%%%%%%%%%%%%%%%%%%%%%%%%%%%%%%%%%%%%%%%%%%%%%

\section{Summary, and a Look Forward} \label{conclusions}

We have presented a detailed analysis of a complex and rich {\it HST} data set 
from Cycle~17 GO program 11677.  Over the course of 9 months in 2010, this program 
spent 121 primary orbits imaging a field with the ACS instrument, and 121 parallel orbits 
imaging multiple fields with both the UVIS and IR cameras of the newly installed 
WFC3 instrument.  These observations were obtained to address science questions that 
require measuring the complete stellar populations of 47~Tuc, from the faintest hydrogen 
burning dwarfs, through the main-sequence and giant branches, to the coolest and 
faintest white dwarf remnants.

We describe the observational design of this program and analyze over 700 full frame 
{\it HST} images using conventional techniques.  Our analysis used the MultiDrizzle 
image analysis package to produce co-added images that are well sampled and free 
of cosmetic defects.  These {\it single} images in each filter were then subjected to 
iterative PSF-fitted photometry using the DAOPHOT and ALLSTAR programs to yield the 
final photometric, astrometric, and morphological catalogs.   The stack-based 
approach presented in this paper for finding and measuring the faintest stars in 
the field is fundamentally different from alternative approaches that analyze the 
raw pixel data in all the images simultaneously.  The other analysis approach is 
underway and will be presented in an upcoming paper (Anderson et~al.\ 2012).  The 
final catalogs and processed images of all fields will be publically released in 
the near future, upon completion of the science program. 

Through our experimentation, we found that three of the most critical steps that affect the 
quality of the output drizzled image was the accuracy of the input image registration, 
the calculation of the sky offsets between all input images, and the final balance achieved 
between the pixfrac and scale parameters.  Currently, the {\it HST} pipeline only produces 
drizzled images for input exposures obtained within the same visit (typically those that 
use the same guide stars), without any empirical corrections for incidental image shifts, 
rotations, or scale changes.  For pipeline processing of crowded field images, a more careful 
consideration of the background variations in the sky level is also needed.  Accurately 
measuring these offsets leads to a smaller standard deviation in the input images and 
better sigma clipping of deviant pixels.  The pipeline reductions also only produce images 
at a pixfrac of unity and with no rescaling, leading to a gross loss of resolution.  Based on our 
experience, the quality of such images can be greatly improved by using the full capabilities of 
MultiDrizzle in post-processing, and so several of the key steps are explained in detail in 
Sections~\ref{MDACS} and \ref{MDWFC3} (see also ``The MultiDrizzle Handbook'' -- Fruchter \& 
Sosey 2009).  Some of this discussion is especially valuable to 
WFC3 users, as this program likely represents the most comprehensive test of the image-analysis 
software on data from the new instrument (e.g., the 59 orbit IR Stare field).  

Related to photometric methods, first, we found that simple (small) aperture photometry on 
the final drizzled images yields an excellent CMD that is not grossly different from that 
achieved with careful PSF-fitting with DAOPHOT.  Our method does improve such first order 
photometry by building a PSF from isolated, bright stars on the frame, where the input 
catalogs were screened several times to remove objects with anomalous measurements.  This 
method has the advantage over techniques where a temporally-varying PSF is derived for each 
of the individual images in that only a single average PSF is measured on the final stack.  
A key disadvantage in some 
implementations of working with drizzled images is that the degree of spatial variation mapped 
across the field may not be appropriate for {\it HST}, for example, if not enough stars are 
present in the field.  In the drizzled images of this program, much of the fine-scale variation 
of the PSF is averaged out through the combination of many images with individual PSFs at 
different orientations and detector locations.

The final photometric catalogs contain over 70,000 stars in the ACS primary field and over 
50,000 stars in the WFC3 Swath fields. The CMDs from the single drizzled images exhibit 
tightly defined sequences of the 47~Tuc main-sequence, background SMC main-sequence, and 
the 47~Tuc white dwarf cooling sequence.  The photometry in the primary ACS field extends 
to $>$30th magnitude in $F606W$ and reveals both very low-mass hydrogen-burning stars in 
the cluster and very cool white dwarf remnants.  The parallel WFC3/UVIS data, with 
exposure times of just 1 orbit per filter per field, exhibit a similar CMD showing the three 
populations well separated.  The sensitivity of the WFC3/UVIS camera is excellent 
in these relatively crowded regimes, reaching depths of $F390W$ = 28.5 and $F606W$ = 29.0 
in a single orbit of integration.\footnote{The $F606W$ photometry shown in Figure~\ref{fig:WFC3CMD} 
includes only those objects that were also detected in the shallower $F390W$ images.}

The WFC3/IR observations in this program represent the deepest IR probe of a resolved 
stellar population to date.  Very low mass 47 Tuc stars are measured over 60 arcmin$^2$ 
with just 3 orbits of integration per field (1 orbit in $F110W$ and 2 orbits in $F160W$).  
The stellar main-sequence is clearly seen to $F110W$ $\gtrsim$ 23, and the photometry 
extends several magnitudes below this.  The morphology of the CMD shows a distinct ``kink'' on the 
lower main sequence, below which the sequence becomes {\it bluer} in $F110W - F160W$ with lower 
luminosity.  The kink is caused by an opacity effect in the atmospheres of cool stars that 
develop molecular hydrogen at $T_{\rm eff}$ = 4,300~K.  This feature will provide added 
leverage to a range of studies where CMDs are modeled to constrain stellar evolution 
and structure theory, as well as cool atmosphere models.

The WFC3/IR CMD of 47~Tuc also provides encouragement for studies of resolved stellar 
populations with the James Webb Space Telescope (JWST).  The NIRCam (and NIRISS) instruments 
on JWST will be sensitive to optical and IR wavelengths extending from 0.6 -- 5.0 microns.  
Through both a short and long wavelength channel, NIRCam will also be diffraction limited at both 
2 and 4 microns.  Relative to HST WFC3/IR, the instrument will have smaller pixels by a factor of 4, 
a larger field of view by more than a factor of two, and much higher overall total system 
throughput.  For example, in $\sim$10,000~seconds of exposure time, the complete stellar populations 
in dense star clusters down to the hydrogen burning limit will be easily measured out to systems 
beyond 50~kpc in the Milky Way (e.g., $J$ = 30th magnitude).

%0.1 Msun H burning limit at about F110W = 11 --> 11+13.3 = 24.3 in 47 Tuc @4500 pc
%NIRCam M5V normalized to J = 27.55 in 1000s reaches S/N = 5 in F150W, but M5V is too hot for H burn limit.
%NIRCam bb = 2500 normalized to J = 29.9 in 10000s reaches S/N = 3 in F200W.

%WFC3/IR difficulties with undersampled data, in building PSF, requires good dithers.  
%But, aperture photometry reveals impressive CMD.  Dithering is not needed for CRs.
%FUTURE - Artificial Star Tests, Proper Motions, Photometry on Individual Images and Comparison

%MAKE PLOT OF A RADIAL CURVE OF A FEW STARS SHOWING ON FLT FRAMES (UNDERSAMPLED - 
%ALL FLUX IN CENTRAL PIXEL) AND ONE WITH 2.5 FWHM.

%%%%%%%%%%%%%%%%%%%%%%%%%%%%%%%%%%%%%%%%%%%%%%%%%%%%%%%%%%%%%%%%%%%%%%%%%%%%%%%%%%%%

\acknowledgements
We wish to thank Anton Koekemoer and Andy Fruchter for numerous discussions related 
to the use of MultiDrizzle.  We wish to thank St\'{e}phane Guisard for providing us with his 
beautiful wide-field ground based image of the SMC and 47~Tuc.  Support for program GO-11677 
was provided by NASA through a grant from the Space Telescope Science Institute, which is 
operated by the Association of Universities for Research in Astronomy, Inc., under NASA 
contract NAS 5-26555.  JSK was supported for this work through an STScI Director's 
Discretionary Fund grant.  HBR is supported by grants from The Natural Sciences and
Engineering Research Council of Canada and by the University of British Columbia.

%%%%%%%%%%%%%%%%%%%%%%%%%%%%%%%%%%%%%%%%%%%%%%%%%%%%%%%%%%%%%%%%%%%%%%%%%%%%%%%%%%%%

\appendix
\renewcommand\thesection{A\Alph{section}}

The ACS data set presented in this paper is unlike most {\it HST} investigations of resolved stellar 
populations.  Many more images of 47~Tuc were collected in our program than are typically available, and 
this provides exquisite screening of deviant pixels and leads to clean, ultra-deep stacks of the data.  
To test the specific methods described above on more typical data sets, we re-analyzed {\it HST}/ACS 
archive observations of the rich globular cluster NGC~2808 from MAST (GO-10922; PI.\ G.\ Piotto).  These 
observations consist of just four dithered exposures in $F475W$ and six dithered exposures in $F814W$, 
each with exposure times of 350 -- 360~s.  A single short observation was also obtained in each filter 
for the brighter stars.  

\begin{figure*}[ht]
\begin{center}
\leavevmode 
\includegraphics[width=12.0cm, angle=270]{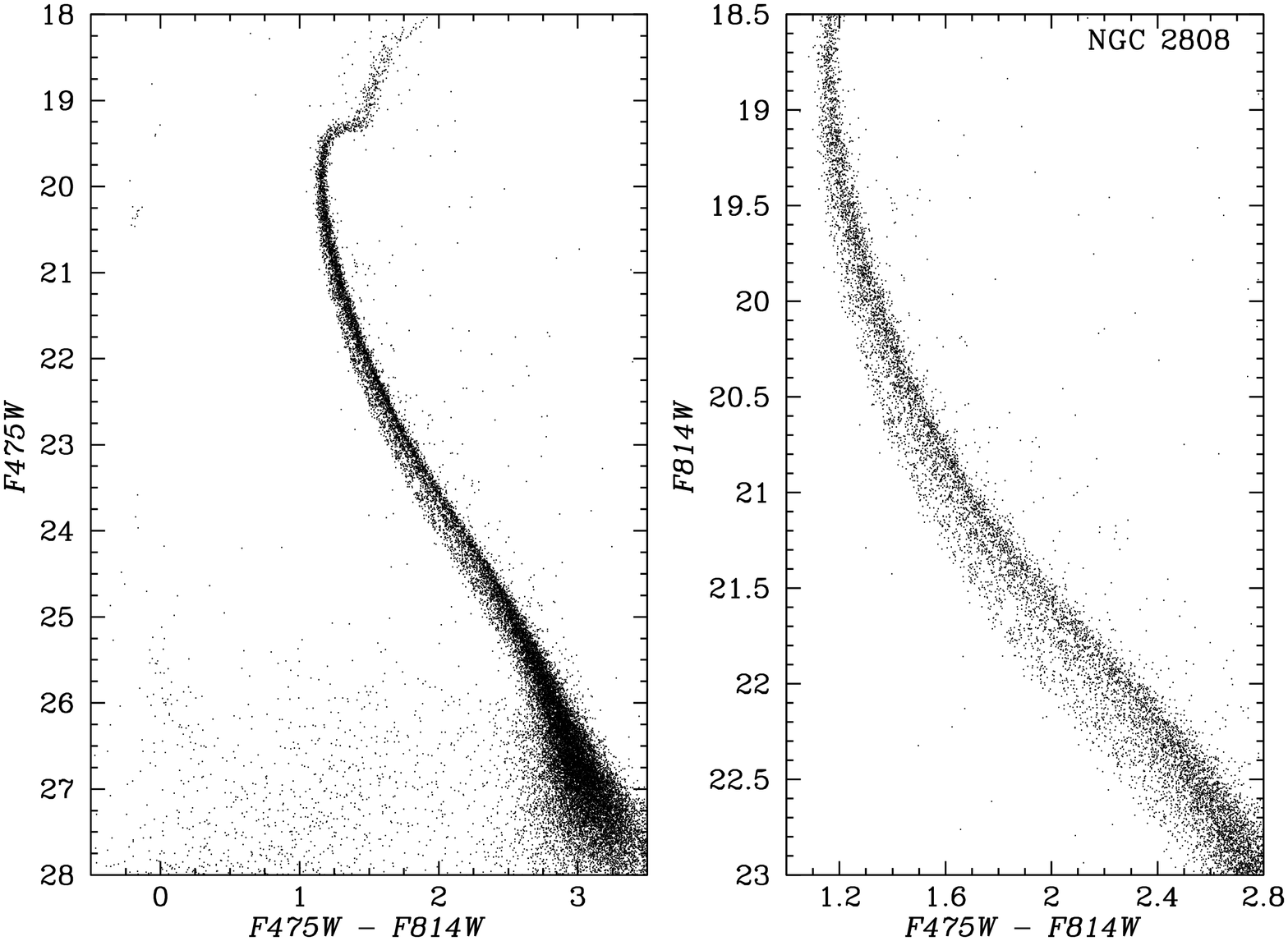}
\end{center}
\vspace{-0.5cm}
\caption{The specific methods described in this paper to register images, subtract sky, combine 
distortion free frames, and perform iterative PSF-fitted photometry are applied to the GO-10922 
{\it HST}/ACS archive observations of NGC~2808 (see Piotto et~al.\ 2007).  These observations are 
very different from our 47~Tuc ACS data, and consist of just a handful of exposures in each of the two 
filters, with exposure times of 350 -- 360~s.  Through a fully automated 
analysis, we measure the main-sequence of the cluster down to 28th magnitude and a white dwarf 
cooling sequence.  A closer look at the cluster's main-sequence shows the three main-sequences 
reported by Piotto et~al.\ (2007).  These data have not been proper motion cleaned and have not been 
corrected for differential reddening. \label{fig:NGC2808}}
\end{figure*}

The results from the analysis of the GO-10922 data are fully described in \cite{piotto07}.  From careful 
PSF-fitted photometry using the methods of Anderson et~al.\ , Piotto et~al.\ are able to resolve NGC~2808's 
main-sequence into three distinct sequences.  Our independent analysis of these data is shown in 
Figure~\ref{fig:NGC2808}.  This includes iterative image registration and sky subtraction, stacking with 
MultiDrizzle, and PSF-fitted photometry with DAOPHOT and ALLSTAR.  All of the steps were done in an automated 
way, and the total analysis took $<$10 hours from the time the data were retrieved from the archive to the 
final zero-point calibrated CMD.

The cluster main-sequence on the CMD in Figure~A1 extends down to $\sim$28th magnitude and shows a 
rich main-sequence with multiple splittings, converging at the turnoff.  A white dwarf cooling sequence is also 
revealed in the faint-blue part of the CMD, previously not reported.  The panel on the right shows a zoomed region 
of the CMD focused on the main sequence, and illustrates the three distinct splittings.  Unlike the Piotto et~al.\ 
analysis, we have not proper motion cleaned these data and have not corrected for differential reddening effects, both of 
which would make the delineation of the sequences stronger.  This simple test confirms that the methods described 
in this paper, when properly implemented, can yield high precision photometry in crowded regimes for data sets with 
a handful of exposures collected over 1 -- 2 orbits.

%%%%%%%%%%%%%%%%%%%%%%%%%%%%%%%%%%%%%%%%%%%%%%%%%%%%%%%%%%%%%%%%%
%%%%%%%%%%%%%%%%%%%%%%%%%%%%%%%%%%%%%%%%%%%%%%%%%%%%%%%%%%%%%%%%%

%%%%%%%%%%%%%%%%%%%%%%%%%%%%%%%%%%%%%%%%%%%%%%%%%%%%%%%%%%%%%%%%%
%%%%%%%%%%%%%%%%%%%%%%%%%%%%%%%%%%%%%%%%%%%%%%%%%%%%%%%%%%%%%%%%%

%%%%%%%%%%%%%%%%%%%%%%%%%%%%%%%%%%%%%%%%%%%%%%%%%%%%%%%%%%%%%%%%%%%%%%%%%%%%%%%%%%%%

\end{document}